# Solving the Large Scale Next Release Problem with a Backbone Based Multilevel Algorithm

Jifeng Xuan, He Jiang, *Member*, *IEEE*, Zhilei Ren, and Zhongxuan Luo

**Abstract**—The Next Release Problem (NRP) aims to optimize customer profits and requirements selection for the software releases. The research on the NRP is restricted by the growing scale of requirements. In this paper, we propose a Backbone based Multilevel Algorithm (BMA) to address the large scale NRP. In contrast to direct solving approaches, BMA employs multilevel reductions to downgrade the problem scale and multilevel refinements to construct the final optimal set of customers. In both reductions and refinements, the backbone is built to fix the common part of the optimal customers. Since it is intractable to extract the backbone in practice, the approximate backbone is employed for the instance reduction while the soft backbone is proposed to augment the backbone application. In the experiments, to cope with the lack of open large requirements databases, we propose a method to extract instances from open bug repositories. Experimental results on 15 classic instances and 24 realistic instances demonstrate that BMA can achieve better solutions on the large scale NRP instances than direct solving approaches. Our work provides a reduction approach for solving large scale problems in search based requirements engineering.

## 1 INTRODUCTION

For a large software project, determining the requirements assignment in the next release is an important problem in the requirements phase [12]. The customers of the software wish to purchase the products suitable for their needs while the software company wishes to select optimal requirements to maximize commercial profits [48]. Due to the complexity of customer requests and product requirements, decisions for software releases frequently conflict with efforts to maximize the profits of the project [10]. To maximize the profits of a software project, an ideal approach is to implement all the requirements to satisfy each potential customer. Limited by the software costs (e.g., the budget or the development time), only a subset of these requirements can be selected in the next release [48]. From the perspective of the software company, the goal of the next release is to select the optimal requirements to maximize the customer profits. However, two factors restrict the development of requirements selection: the problem scale and the requirements dependency.

On one hand, when facing large scale requirements management, it is time-consuming to optimize customer profits [49]. The growth in scale has been listed as one of the nine research hotspots in the future of requirements engineering [10]. Some studies have focused on the large scale requirements analysis. For example, 1000 requirements are provided for the experiments on elicitation and triage in the *SugarCRM* project [9], [14]; 2400 market requirements and 1100 business requirements are handled for the next release in the *Baan* software framework [48]. Although some optimization technologies are introduced to balance the customer profits and requirements costs, such as the cost-value approaches [32], the linguistic-engineering approaches [48], [49], and the search based approaches [5], [57], it is still a challenge to select an optimal decision for large scale requirements problems [67].

On the other hand, the requirements dependencies increase the complexity of requirements optimization. In the modern incremental software development process, new-coming requirements may build joint functions with previous and associate requirements [50]. An industry survey shows that about 80% requirements are constrained by dependencies, which significantly complicate the decision for the software releases [8].

In this paper, we address large scale requirements selection with the Next Release Problem (NRP), which is proposed to model the decision for customer profits and requirements costs [4]. The NRP seeks to maximize customer profits from a set of dependent requirements, under the constraint of a predefined budget bound. Assisted by the NRP, a requirements engineer can make a decision for software requirements to balance the profits of the company and the customers. As a combinatorial optimization problem, the NRP has been proved as "$\mathcal{NP}$-*hard* even when it is basic and customer requirements are independent" [4], i.e., unless $\mathcal{P} = \mathcal{NP}$, there exists no exact algorithm to select the optimal set of customers to maximize the profits in polynomial time [18]. In practice, especially for a large scale problem, it is hard to exactly optimize the decision of the NRP. Thus, it is straightforward to design approximate algorithms to generate near-

- *J. Xuan and Z. Ren are with the School of Software, Dalian University of Technology, Dalian 116621, China. E-mail: {xuan, ren}@mail.dlut.edu.cn.*
- *H. Jiang is with the School of Software, Dalian University of Technology, Dalian 116621, China. E-mail: hejiang@ieee.org.*
- *Z. Luo is with the School of Mathematical Sciences, Dalian University of Technology, Dalian 116621, China. E-mail: zxluo@dlut.edu.cn.*

optimal decisions within polynomial time. Many search-based approaches have been proposed to approximately solve the NRP and its variant problems, including greedy algorithms [4], [25], greedy randomized adaptive search procedures [4], local searches (e.g., a hill climbing or a simulated annealing) [4], [5], genetic algorithms [13], [57], ant colony optimizations [31], [13], etc. A few of these approaches (e.g., a simulated annealing [7] and a genetic algorithm [52]) can work effectively on the small scale NRP. However, these approaches for the small scale problems cannot be directly applied to the large scale problems. For example, Natt och Dag et al. [48] show the hardness of large scale requirements management by analyzing the relationship between customer requests and product requirements; Svahnberg et al. discuss the growing complexity while the problem scale improves [59]. For the large scale NRP [1], it is necessary to design an effective algorithm to cope with the increasing problem scale.

In this paper, we propose a Backbone based Multilevel Algorithm (BMA) to iteratively solve the large scale NRP. In contrast to direct solving approaches, BMA iteratively downgrades the scale of the problem by fixing the *backbone*, which can be approximately viewed as the common part of customers with optimal requirements. The backbone is one of the effective tools in large scale combinatorial optimization in recent years [58], [64], [30]. In our work, two kinds of backbones are employed for the NRP, namely the *approximate backbone* (the common part of customers from a given number of local optimal solutions) and the *soft backbone* (the customers, who add zero cost to the requirements selection). Based on the backbone, we can break a large scale problem down to small ones and refine the solution to the original problem.

To face the lack of open large requirements databases, we propose a method to mine the NRP instances[2] from open bug repositories. Knowledge from bug repositories is extracted to generate the information of requirements. For example, we map a bug report and a user in open bug repositories to a requirement and a customer in requirements engineering. Based on our method, we can generate realistic NRP instances from open bug repositories.

In our work, first, we give the definitions of the NRP model and illustrate the NRP with an example. Then, we define the NRP backbone and propose the instance reduction approach for the NRP. In the implementation of the backbone, we use the approximate backbone to replace the backbone and present the similarity between the approximate backbone and the backbone via the fitness landscape analysis; to augment the application of the approximate backbone, we propose the new concept of the soft backbone. Next, we employ the approximate backbone and the soft backbone to design BMA, which employs a multilevel strategy to enhance the instance reduction. The framework of BMA includes three phases, namely reduction, solving, and refinement. Finally, experiments are conducted on 15 instances generated from the classic literature and 24 new instances extracted from open bug repositories. Experimental results show that our BMA can achieve better solutions than direct solving algorithms on large scale NRP instances.

The primary contributions of this paper are as follows:

1. We present a new algorithm, BMA, to reduce the problem scale of the NRP. In this algorithm, we show how to incorporate the backbone into an approximate algorithm for solving large scale problems. To our knowledge, this is the first application of the backbone in requirements engineering.
2. We propose the soft backbone to augment the existing concept of the backbone in both software engineering and combinatorial optimization. In our work, the soft backbone is directly obtained from the instance after the instance reduction by fixing the selected near-optimal customers.
3. We generate new NRP instances from bug repositories of three open source software projects. The bug repositories are mined to cope with the lack of open requirements repositories. This method of mining new instances can provide realistic instances for the empirical research.
4. We experimentally evaluate BMA and two other existing algorithms for large scale NRP instances. Numerous experimental results on both solution quality and running time are shown to present the performance of these algorithms.

This paper is an extension of our previous work presented at Search Based Software Engineering (SBSE) Track at the 12th Annual Conference on Genetic and Evolutionary Computation (GECCO '10) [28]. In this extension, we add the new concept of the soft backbone, the new method for instance generation, and numerous experimental results.

The remainder of this paper is organized as follows. Section 2 states the definitions of the NRP. Section 3 shows the NRP backbone and the instance reduction. In Section 4, we propose BMA for the NRP. Section 5 presents the experiments and results. Section 6 gives the threats to validity in our work. Section 7 lists the related work. Section 8 briefly concludes this paper and presents our future work.

## 2 Problem Definitions

In this section, we present the related definitions of the NRP and illustrate the NRP with an example instance.

The NRP can be retrieved from the following scenario [4]: in the requirements analysis phase of a software project, a necessary step is to select adequate requirements in the next release to achieve maximized commercial profits within a limited cost. Each customer requests a fraction of candidate requirements and provides a potential commercial profit for the software company. In a real-world project, the dependencies among candidate requirements restrict the selection of customer profits. The NRP aims to determine a subset of customers to achieve maximum profits under a predefined budget bound.

[1] Throughout this paper, the *large scale NRP* can be viewed as the NRP in large scale requirements management, which is to balance the customer profits and requirements costs in large scale software projects.

[2] An *instance* is a detailed model generated by specifying particular values for all the parameters of a problem [45].

According to this application scenario, we give the formal definitions of the NRP as follows. In a software project, let $R$ be the set of all the candidate requirements and the cardinality of $R$ is $|R| = m$. Each requirement $r_j \in R$ $(1 \le j \le m)$ is associated with a nonnegative cost $c_j \in C$. A directed acyclic graph $G = (R, E)$ denotes the dependencies among these requirements, where $R$ is the set of vertexes and $E$ is the set of arcs. In the dependency graph $G$, an arc $(r', r) \in E$ indicates that the requirement $r$ depends on $r'$, i.e., if $r$ is implemented in the next release, $r'$ must be implemented as well to satisfy the dependency. The requirement $r$ is called the child requirement of $r'$. Let $parents(r)$ be the set of requirements, which can reach $r$ via one or more arcs. More formally, $parents(r) = \{r' \in R | (r', r) \in E \; or \; (r', r'') \in E, r'' \in parents(r)\}$. All the requirements in $parents(r)$ must be implemented to ensure the implementation of $r$.

Let $S$ be all the customers related to the requirements $R$ and $|S| = n$. Each customer $s_i \in S$, requests a set of requirements $R_i \subseteq R$. Let $w_i \in W$ be the profit gained from the customer $s_i$. Let $parents(R_i) = \cup_{r \in R_i} parents(r)$. For a given customer $s_i$, let the set of total requirements requested by $s_i$ be $\hat{R}_i = R_i \cup parents(R_i)$. Under the above definitions, a customer $s_i$ can be satisfied by the software release decision, if and only if all the requirements in $\hat{R}_i$ are implemented in the next release. Let the cost for satisfying the customer $s_i$ be $cost(\hat{R}_i) = \sum_{r_j \in \hat{R}_i} c_j$. A subset of customers $S_0 \subseteq S$ can be viewed as a solution. To facilitate the following discussion, we also formulate a solution as a set of ordered pairs, i.e., the solution $S_0 \subseteq S$ is denoted as $X = \{(i, p) | p = 1, s_i \in S_0 \; or \; p = 0, s_i \notin S_0\}$. It is easy to convert the form of $X$ or $S_0$ into each other. Let the cost of a solution $X$ be $cost(X) = cost(\cup_{(i,1) \in X} \hat{R}_i)$ and the objective function of $X$ (i.e., the profit of $X$) be $\omega(X) = \sum_{(i,1) \in X} w_i$.

**Definition 1.** *The next release problem (NRP).*

Given a directed acyclic requirements dependency graph $G = (R, E)$, each customer $s_i \in S$ directly requests a set of requirements $R_i$. The profit of $s_i$ is $w_i \in W$ and the cost of requirement $r_j \in R$ is $c_j \in C$. A predefined budget bound is $b$.

The goal of the NRP is to find an optimal solution $X^*$, to maximize $\omega(X)$, subject to $cost(X) \le b$.

For an NRP instance, the scale is $n = |S|$. To simplify the statements, all the values of an NRP instance are integers except especial specifications. For a real-world application, it is easy to convert a non-integer NRP instance into an integer-only instance by magnifying the same multiple for all the values.

An NRP instance with 7 customers and 8 requirements is illustrated as follows. The requirements are extracted from a communication company project, which is introduced in [50]. Table 1 shows the description of these 8 requirements and the dependencies. In Fig. 1, we present the dependency graph and the requirements requested by customers, where the arrows from top to bottom indicate the dependencies and the lines indicate customer requests. For the requirements set $R = \{r_1, r_2, \dots, r_8\}$, let the costs $c_1, c_2, \dots, c_8$ of these requirements be 2, 5, 4, 3, 8, 1, 5, 2, respectively; for the customer set $S = \{s_1, s_2, \dots, s_7\}$, let the profits $w_1, w_2, \dots, w_7$ of these customers be 7, 2, 6, 5, 4, 3, 1. Taken $s_1$ as an example, the total requirements requested by $s_1$ are $\hat{R}_1 = \{r_1, r_2, r_6\}$; the cost for satisfying the requirements is $cost(\hat{R}_1) = 8$; and the profit of $s_1$ is $w_1 = 7$.

TABLE 1
REQUIREMENTS AND DEPENDENCIES

| Requirement | Description | Cost | Arc |
|---|---|---|---|
| $r_1$ | Expanding memory on BTS controller | 2 | |
| $r_2$ | BTS variant | 5 | |
| $r_3$ | Market entry feature 1 | 4 | |
| $r_4$ | Market entry feature 2 | 3 | $(r_3, r_4)$ |
| $r_5$ | Market entry feature 3 | 8 | $(r_4, r_5)$ |
| $r_6$ | Next generation BTS | 1 | $(r_2, r_6)$ |
| $r_7$ | Pole mount packaging | 5 | $(r_2, r_7)$ |
| $r_8$ | Software quality initiative | 2 | |

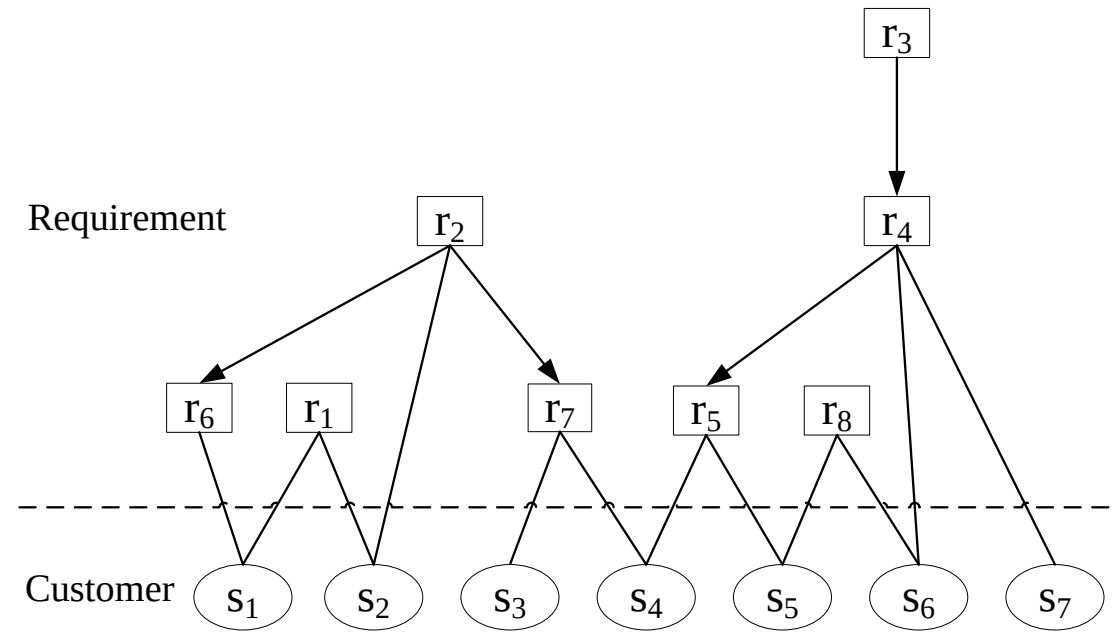

Fig. 1. Requirements dependencies and customer requests.

Given a predefined budget bound $b = 26$, the profit and the cost of a feasible solution $X_1 = \{(1,1), (2,0), (3,1), (4,0), (5,0), (6,0), (7,1)\}$ are 14 and 20, respectively. Similarly, the profit and the cost of $X_2 = \{(1,1), (2,0), (3,0), (4,0), (5,1), (6,1), (7,1)\}$ are 15 and 25. Obviously, $X_2$ is a better solution than $X_1$. However, $X_3 = \{(1,0), (2,1), (3,0), (4,1), (5,0), (6,1), (7,0)\}$ is unfeasible since its cost 29 exceeds the bound $b$.

From the definition of the NRP, the requirements $\hat{R}_i$ requested by a customer $s_i$ are calculated from the dependency graph of requirements [4]. If we directly input the requirements requests for each customer, Definition 1 can be simplified [5], [69].

**Definition 2.** *The Simplified NRP.*

Given a set of requirements $R$ and a set of customers $S$, each requirement $r_j \in R$ $(1 \le j \le m)$ has a cost $c_j \in C$ and each customer $s_i \in S$ $(1 \le i \le n)$ has a profit $w_i \in W$. A request $q_{ij} \in Q$ shows whether a customer $s_i$ requests a requirement $r_j$ in the next release, i.e., $q_{ij} = 1$ denotes that $s_i$ requests $r_j$ or $q_{ij} = 0$ denotes not. Given a solution $X$, the requirements for $X$ is $R(X) = \cup_{(i,1) \in X, q_{ij} = 1} \{r_j\}$. A predefined budget bound is $b$.

The goal of the NRP is to find an optimal solution $X^*$, to maximize $\omega(X) = \sum_{(i,1) \in X} w_i$, subject to $cost(X) = \sum_{r_k \in R(X)} c_k \le b$.

Based on the definitions, each NRP instance can be directly converted into a Simplified NRP instance. The dependencies among requirements are included in the requirements requests $Q$. To simplify the following statements, a Simplified NRP instance is called an NRP instance for short. We denote an NRP instance as $\Pi$.

## 3 BACKBONE BASED INSTANCE REDUCTION

In this paper, for the large scale NRP, our basic idea is to reduce the scale of an NRP instance to get a small instance, which is easy to solve. In this section, we will present the backbone based instance reduction and the substitutions of the backbone, namely the approximate backbone and the soft backbone.

### 3.1 Backbone and Instance Reduction

The backbone is a useful tool for algorithm design in constraint solving and combinatorial optimization [33]. In an algorithm, the backbone is viewed as an ideal structure to model the common characteristics of the optimal solutions [30]. On one hand, if the backbone is ideally obtained, the optimal solutions to an instance can be partly constructed; on the other hand, it is usually intractable to obtain the backbone within polynomial time [58]. In practice, the approximate backbone is usually employed instead. Backbone based algorithms have been shown effective on some classic problems, such as the Maximum SATisfiability (Max-SAT) [64], the Travelling Salesman Problem (TSP) [33], and the Quadratic Assignment Problem (QAP) [30]. If we consider searching for a solution as finding a key part in a physical body, the backbone can be informally viewed as the common part of the global optimal solutions. A *global optimal solution* (called an *optimal solution* for short in the rest of this paper) is defined as the best solution in the whole search space while a *local optimal solution* is the best solution in a local part of the search space with respect to a given algorithm [45]. We define the NRP backbone in Definition 3.

**Definition 3.** *The NRP backbone.*

Given an NRP instance $\Pi$, let $\Gamma^* = \{X_1^*, X_2^*, \dots, X_u^*\}$ be the set of all the optimal solutions to $\Pi$. The backbone of $\Pi$ is defined as $\xi = \bigcap_{i=1}^{u} X_i^*$.

The backbone scale of $\xi$ is $|\xi|$. Based on Definition 3, the NRP backbone contains the common characteristics of the optimal solutions. Given an NRP instance, we can reduce the instance scale by fixing its backbone.

**Definition 4.** *The NRP instance reduction.*

Given an NRP instance $\Pi$ and its backbone $\xi$, an instance reduction is a process to generate a new and small scale instance $\Pi'$, which is easy to solve. Meanwhile, the backbone and a solution to the new instance can be used to form the solution to the original instance.

A new instance $\Pi'$ can be constructed by removing the customers and the requirements of the backbone from the original instance. We list the parameter values for the variables of the new instance in Table 2. For an instance $\Pi$, the customers of its backbone $\xi$ is $S(\xi)$ while the requirements selected in $\xi$ is $R(\xi)$. The request set $Q(\xi)$ indicates the requirements requests of $S(\xi)$. In the new instance $\Pi'$, the customers in $S(\xi)$ and the requirements selected by $S(\xi)$ is not helpful yet. Thus, $S'$, $R'$, and $Q'$ in $\Pi'$ are generated by removing the elements in $S(\xi)$, $R(\xi)$, and $Q(\xi)$, respectively. To build the budget bound $b'$ for $\Pi'$, we remove the cost of $\xi$, i.e., $cost(\xi)$. The profits $W'$ and the costs $C'$ are the subsets related to $S'$ and $R'$. We can validate that each optimal solution to $\Pi$ can be constructed from an optimal solution to $\Pi'$ and the backbone $\xi$. According to Table 2, each parameter value of $\Pi'$ can be obtained within polynomial time. Based on these parameter values, the new instance $\Pi'$ can be uniquely determined, denoted as $\Pi' = \text{Instance-Reduction}(\Pi, \xi)$.

TABLE 2
PARAMETER VALUES FOR THE INSTANCE REDUCTION

| Instance $\Pi$ | Backbone $\xi$ | New instance $\Pi'$ |
|---|---|---|
| $S$ | $S(\xi) = \{s_i \| (i,1) \in \xi \text{ or } (i,0) \in \xi\}$ | $S' = S \setminus S(\xi)$ |
| $R$ | $R(\xi) = \{r_j \| (i,1) \in \xi, q_{ij} = 1\}$ | $R' = R \setminus R(\xi)$ |
| $Q$ | $Q(\xi) = \{q_{ij} \| s_i \in S(\xi) \text{ or } r_j \in R(\xi)\}$ | $Q' = Q \setminus Q(\xi)$ |
| $W$ | | $W' = \{w_i \| s_i \in S'\}$ |
| $C$ | | $C' = \{c_j \| r_j \in R'\}$ |
| $b$ | $cost(\xi) = \sum_{r_k \in R(\xi)} c_k$ | $b' = b - cost(\xi)$ |

However, the NRP backbone is obtained from the optimal solutions, which cannot be obtained within polynomial time. Thus, there exists no polynomial time algorithm to find the NRP backbone. In this paper, we use the approximate backbone and the soft backbone to replace the backbone.

The approximate backbone is the set of common customers of a given number of local optimal solutions while the soft backbone is the set of optimal customers with no cost for the given instance. In Fig. 2, we summarize the relationship among the backbone, the approximate backbone, and the soft backbone. The approximate backbone is generated as the common part of a group of local optimal solutions; the soft backbone is directly extracted from the current instance. The union of the approximate backbone and the soft backbone is called the *combined backbone*. We employ the combined backbone to build the near-optimal solutions for the NRP. In Section 3.2 and Section 3.3, we will present more details about the approximate backbone and the soft backbone, respectively.

### 3.2 Approximate Backbone

Since no polynomial time algorithm exists to exactly obtain the NRP backbone, we follow the existing work to replace the backbone with the approximate backbone, which is generated from a set of local optimal solutions [64]. In this section, we will present the relationship between the optimal solutions and local optimal solutions using the fitness landscape analysis. We show that the approximate

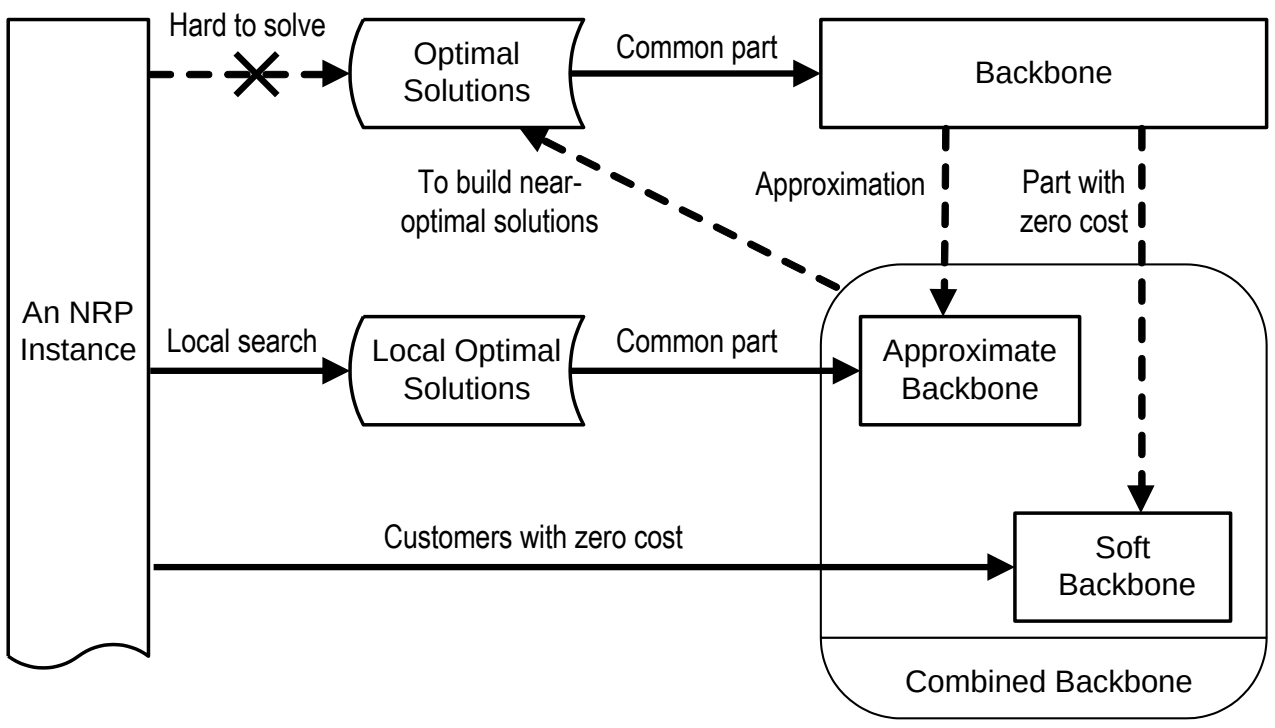

Fig. 2. The relationship among the backbone, the approximate backbone, and the soft backbone.

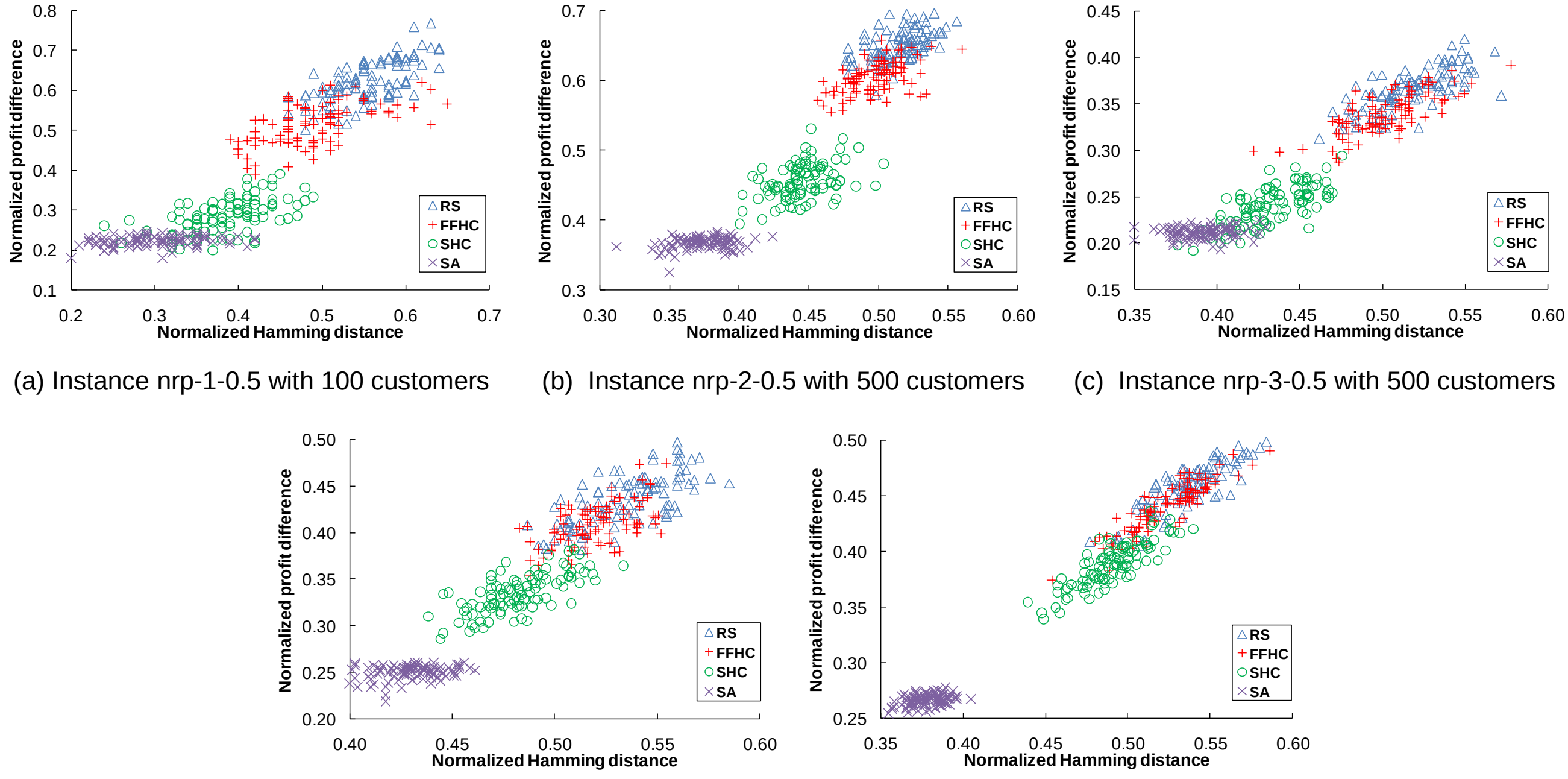

(a) Instance nrp-1-0.5 with 100 customers (b) Instance nrp-2-0.5 with 500 customers (c) Instance nrp-3-0.5 with 500 customers

(d) Instance nrp-4-0.5 with 750 customers (e) Instance nrp-5-0.5 with 1000 customers

Fig. 3. Landscape of five classic NRP instances with four algorithms. For each instance in a sub-figure, the x-axis is the normalized Hamming distance from a local optimal solution to the optimal solution and the y-axis is the normalized profit difference of these two solutions. In each sub-figure, the point (0,0) denotes the optimal solution. A solution in the bottom-left corner is more similar to the optimal solution than that in the top-right corner.

backbone can partly reflect the characteristics of the backbone.

Compared to the concept of backbone, an approximate backbone of an NRP instance is defined as the common part of local optimal solutions. In real-world applications, a local optimal solution is generated within polynomial time by a local search algorithm, which is also called as a *local search operator* when it is incorporated into another algorithm [45]. We give Definition 5 to describe the approximate backbone.

**Definition 5.** *The NRP approximate backbone.*

Given an NRP instance $\Pi$, let $\Gamma = \{X_1, X_2, \dots, X_v\}$ be a set of local optimal solutions to $\Pi$. The approximate backbone of $\Pi$ is defined as $\xi^a = \text{Approximate-Backbone}(\Pi, \Gamma) = \bigcap_{i=1}^{v} X_i$.

We employ the fitness landscape analysis [44] to investigate the relationship between the backbone and the approximate backbone. The fitness landscape analysis is an important technology to understand the behavior of combinatorial optimization algorithms [44]. For large scale optimization, the fitness degree and the solution distribution in the fitness landscape are measured to guide the design of algorithms [64]. To analyze the fitness landscape between the backbone and the approximate backbone, we evaluate the differences between the optimal solutions and local optimal solutions by the distances of these two kinds of solutions. The distance is usually measured as Hamming distance [44], [54]. For an NRP instance with scale $n$, the Hamming distance between a solution $X$ and an optimal solution $X^*$ is $dist(X, X^*) = n - |X \cap X^*|$. Thus, the normalized Hamming distance is $n\text{-}dist(X, X^*) = (n - |X \cap X^*|)/n$. To evaluate the difference of profits between these two solutions, we define the normalized profit difference as $n\text{-}diff(X, X^*) = \big(\omega(X^*) - \omega(X)\big)/\omega(X^*)$. In practice, the optimal solutions to large scale instances are hard to obtain. Therefore, we follow the existing fitness landscape analysis approaches to replace the optimal solutions with the best known solutions[3] [44], [54]. Note that the measure criterion of the profit difference in our work is a little different from that in some existing work on the fitness landscape (e.g., [42], [43]). In our work, we use the normalized profit differences between solutions to evaluate the relationship between the backbone and the approximate backbone while the existing work uses the fitness values of solutions to evaluate the fitness degree.

In Fig. 3, we present the fitness landscape of five classic NRP instances. The scales of these instances are 100, 500, 500, 750, and 1000, respectively (see Section 5.2 for the details of these instances). In the fitness landscape analysis, we employ four algorithms to indicate the similarity between local optimal solutions and the optimal solutions, including Randomized Search (RS), First Found Hill Climbing (FFHC), Sampled Hill Climbing (SHC), and Simulated Annealing (SA). RS is a randomized algorithm, which randomly generates a solution. Due to the budget bound of the NRP, a solution may be infeasible; RS repairs these infeasible solutions by randomly removing a couple of selected customers. FFHC and SHC are two kinds of hill climbing algorithms proposed in [4]. As their names suggest, FFHC updates its solution with the first improved solution while SHC updates its solution with the best solution among a certain number of sampled so-lutions. SA (also called LMSA in [4]) is an extension of a non-linear simulated annealing algorithm, which combines the hill climbing with an acceptance temperature [4], [5]. Among these four algorithms, SA usually obtains the best solutions for the NRP [4]. In the experiments, each of these four algorithms is independently run for 100 times and each fitness landscape of an algorithm consists of 100 local optimal solutions.

[3] To obtain a best known solution of an NRP instance, RS (in Section 3.2) and BMA (in Section 4.2) have been performed repeatedly ($10^5$ times for RS and 200 times for BMA, respectively) and the best solution is selected.

As shown in Fig. 3, an algorithm with small distances between solutions, can provide small profit differences. For FFHC, SHC, or SA, the distances between solutions are from 0.2 to 0.6 of the instance scale while the profit differences between solutions are from 0.2 to 0.7 of the profits of the best known solutions. In general, when measuring the distances for these five instances, SA is better than SHC and SHC is better than FFHC. On most of the five instances, SA is the best algorithm, which leads to both small distances and small profit differences. An exception instance is nrp-1-0.5, the one with the smallest scale. On nrp-1-0.5, the landscape of SHC covers the landscape of SA. This fact is primarily due to the small scale of nrp-1-0.5, i.e., both SA and SHC can generate good solutions. Moreover, for all the instances, the distances by SA are less than 0.45 of scales. Compared with FFHC and SHC, both the solution distances and profit differences by SA are stable. The other algorithm, RS, only provides large distances around 0.6 of instance scales.

The fitness landscape analysis indicates that there is an overlap between local optimal solutions and the optimal solutions. Thus, local optimal solutions can partly represent the characteristics of the optimal solutions. Meanwhile, among these four algorithms, an algorithm with high performance leads to small solution distances. Thus, a local optimal solution by a high-performance algorithm can do well in showing the characteristics of the optimal solutions. In summary of the fitness landscape analysis, the NRP backbone can be replaced by the approximate backbone, which is the intersection of local optimal solutions obtained by a good local search algorithm.

The instance reduction in Definition 4 can be applied to the approximate backbone. Since local optimal solutions are obtained in polynomial time, the approximate backbone based instance reduction can be conducted within polynomial time.

### 3.3 Soft Backbone

In this section, apart from the approximate backbone, we propose the soft backbone to augment the application of the backbone.

Given an approximate backbone, we can generate a new and small instance after the instance reduction. The new instance can also be viewed as an NRP instance, which can be solved by an existing algorithm. However, there is one key difference between the new instance and the original one. Given an original NRP instance $\Pi$, for each customer $s_i \in S$, $\sum_{r_j \in R} q_{ij} \geq 1$, i.e., $s_i$ requests one or more requirements to be implemented in the next release. Thus, the cost of requirements requested by each customer is more than zero. However, for the new NRP instance $\Pi'$ after an instance reduction, it is possible to find a customer $s_i$ such that $\sum_{r_j \in R'} q_{ij} = 0$. In other words, there may exist a customer, whose requirements have been wholly reduced in the instance reduction. For the goal of the NRP, we add this kind of customer to the solution to maximize the profits. From the perspective of the problem solving, the customers, who provide profits with zero cost, can also be approximately considered as the common part of the optimal solutions.

We define the soft backbone as such customers, who provide profits and request no requirements. In contrast to obtaining the approximate backbone from solutions, the soft backbone is a new kind of backbone obtained from instances.

**Definition 6.** *The NRP soft backbone.*

Given an NRP instance $\Pi'$ after the instance reduction, the soft backbone is defined as $\xi^s$ = Soft-Backbone$(\Pi', \phi)$ $= \{(i,1) | \sum_{r_j \in R'} q_{ij} = 0\}$ ($\phi$ denotes an empty set and $R'$ denotes the requirements set of $\Pi'$).

The instance reduction in Definition 4 can be directly applied to the soft backbone. Thus, for an NRP instance, two instance reductions are employed to reduce the scale of the NRP instance, based on the approximate backbone and the soft backbone, respectively.

In Fig. 4, we take the instance in Fig. 1 as an example to illustrate the instance reductions. Given the approximate backbone $\xi^a = \{(1,1)\}$, the three requirements, $r_1$, $r_2$, and $r_6$, requested by the customer $s_1$ are selected by $\xi^a$. Then the approximate backbone based instance reduction is built and the three requirements and the customer $s_1$ are removed. For the new instance, no requirement is requested by the customer $s_2$. Since $s_2$ provides a profit without any cost of requirements, the soft backbone of the new instance is $\xi^s = \{(2,1)\}$. Based on $\xi^s$, a soft backbone based instance reduction is built and the customer $s_2$ can be removed after this instance reduction. In summary, 2 customers and 3 requirements are removed based on these two instance reductions.

We present four differences between the approximate backbone and the soft backbone in Table 3. First, from the definition, the approximate backbone is the intersection of a given number of local optimal solutions while the soft backbone is directly extracted from the instance. No local search algorithm is needed for obtaining the soft backbone. Second, based on the first difference, the soft back-bone only exists in the instance generated after an instance reduction. Since an original NRP instance does not include the customers who request no requirements, the original instance cannot provide any soft backbone. The soft backbone is a by-product of the instance reduction. In other words, the instance reduction provides an application scenario for the soft backbone. However, the approximate backbone can be generated for all the instances, which have feasible solutions. Third, given a new instance after the instance reduction, the approximate backbone is a kind of approximation of the backbone while the soft backbone is a part of the backbone of this new instance. Based on the definition, the soft backbone can be added to any feasible solution to improve the profit of this solution. Since the optimal solutions have the maximum profit, each optimal solution of the given instance must include the soft backbone. Fourth, only when a customer is selected, this customer may appear in the soft backbone while both selected customers and unselected customers may appear in the approximate backbone.

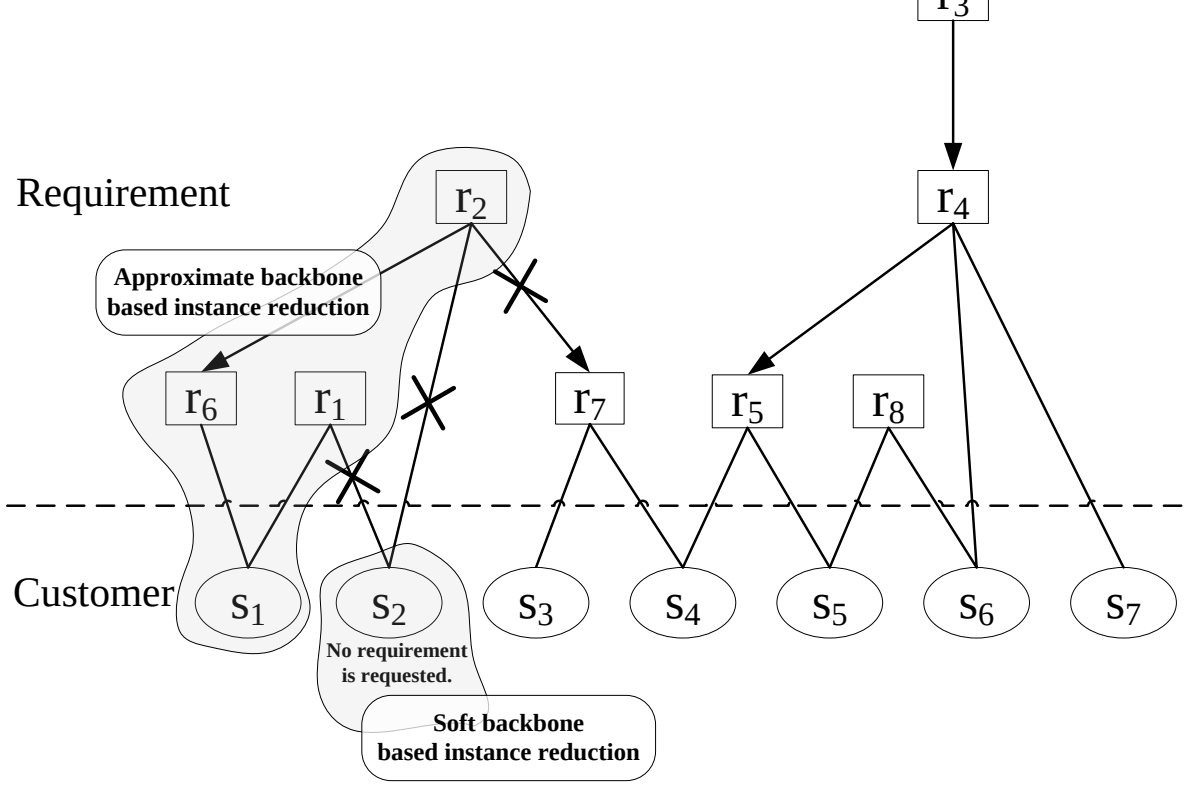

Fig. 4. An example of the approximate backbone and the soft backbone based instance reductions.

TABLE 3
DIFFERENCES BETWEEN THE APPROXIMATE BACKBONE AND THE SOFT BACKBONE

| | Approximate backbone | Soft backbone |
|---|---|---|
| Source | From local optimal solutions | Directly from the instance |
| Existence | Existing for each instance with feasible solutions | Only existing in the new instance after an instance reduction |
| Approximation | Approximation of the backbone | Part of the backbone for the current instance |
| Component | Including both the ordered pairs as $(i_a, 1)$ and $(i_b, 0)$ | Only including the ordered pairs as $(i_a, 1)$ |

## 4 BACKBONE BASED MULTILEVEL ALGORITHM

To address the large scale NRP, we tend to reduce the scale of the NRP instances by fixing the backbone in order to solve the problem with the existing search based algorithms. First, we will show that the multilevel strategy can be employed to iteratively reduce the instance scale. Then we will propose the framework of BMA and illustrate the process of BMA with an example.

### 4.1 Multilevel Strategy

From Section 3.3, we can reduce the scale of an NRP instance using two instance reductions, based on the approximate backbone and the soft backbone, respectively. However, for a large scale instance, the instance after two instance reductions may be still hard to solve with the existing algorithms. Thus, we consider using the multilevel strategy to perform the instance reductions step by step.

A multilevel strategy is to convert the original problem into multiple levels of sub-problems, each of which is an independent problem [60]. In combinatorial optimization, a multilevel strategy includes two kernel phases, namely reduction (reducing the hardness of the problem) and refinement (constructing the solution to the original problem) [60]. In our work, since a new generated instance after one instance reduction can be viewed as a new NRP instance, we use the multilevel strategy to iteratively reduce the scale of an instance, i.e., the original NPR instance is handled with multiple instance reductions and the final solution to the instance is then constructed from the approximate backbones and the soft backbones. In this paper, the approximate backbone and the soft backbone based instance reductions are alternatively used. More specifically, given an instance, we always conduct a soft backbone based instance reduction after an approximate backbone based instance reduction. We call these two instance reductions (based on the approximate backbone and the soft backbone) *a pair of instance reductions* for simplicity.

In Fig. 5, we present the experimental result of the relationship between the pairs of instance reductions and the scales of instances. The instances in Fig. 5 are the same as those in Fig. 3, except the instance nrp-1-0.5 (nrp-1-0.5 is omitted due to its small scale, 100). In this experiment, each approximate backbone is calculated from 5 local optimal solutions, which are obtained by the classic algorithm, SA (see Section 3.2 for details). For each instance, 12 pairs of instances reductions are sequentially used to obtain new and small instances.

As shown in Fig. 5, although a single pair of instance reductions can reduce the scales of instances, it is feasible to employ further reductions when utilizing the multilevel strategy. For example, in the instance nrp-3-0.5 with the scale 500, the scales of two new instances after one pair of instance reductions are 439 and 432, respectively.

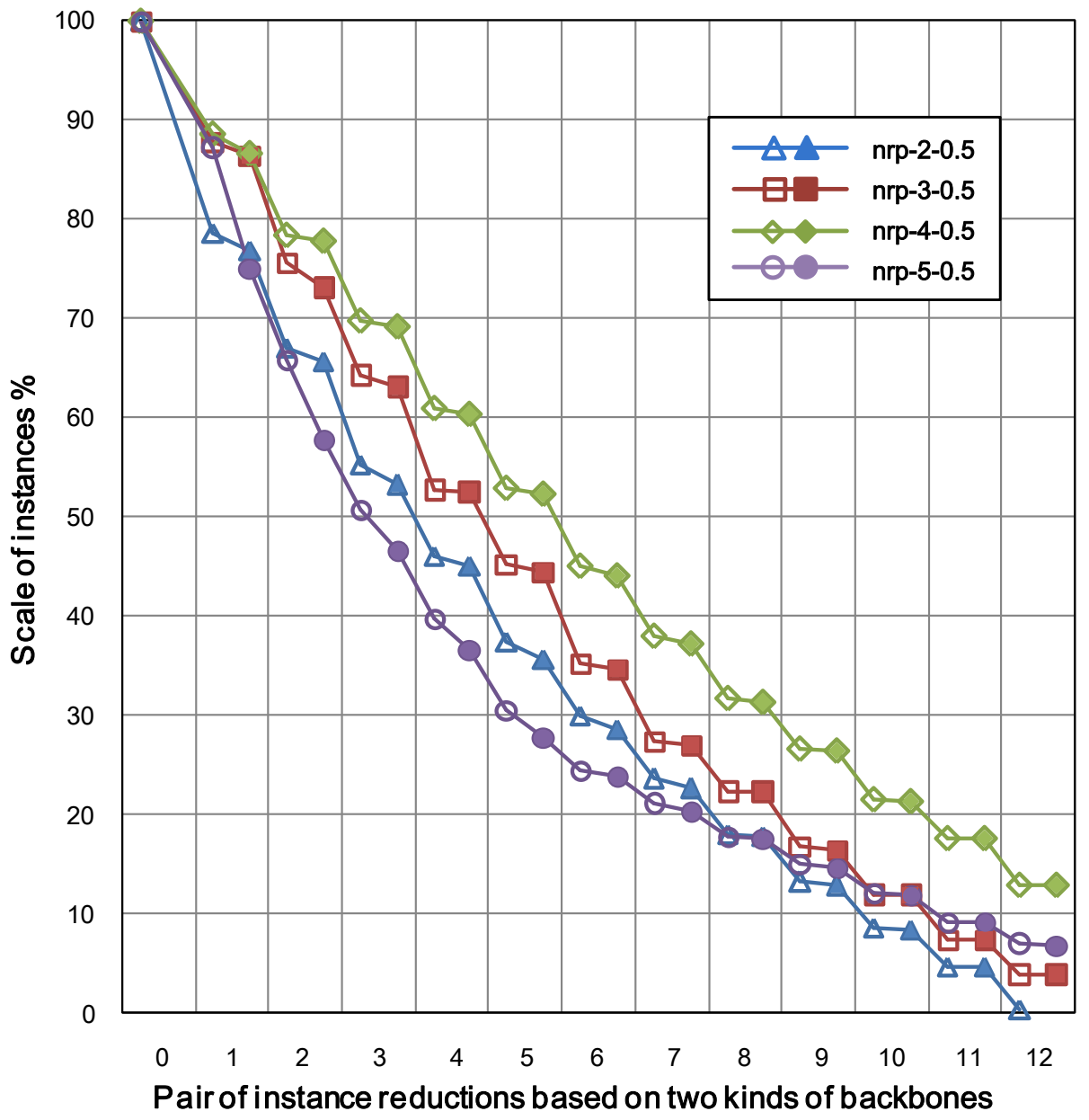

Fig. 5. The instance scales with 12 instance reductions for the approximate backbone and the soft backbone. The scales of the four instances are 500, 500, 750, and 1000. The x-axis shows the pair of instance reductions based on two kinds of backbones and the y-axis shows the change of instance scales. There are two kinds of points in each curve. A solid point denotes an instance reduction based on the soft backbone while the other kind of point denotes an instance reduction based on the approximate backbone.

TABLE 4
TOTAL SCALES REDUCED BY 12 INSTANCE REDUCTIONS

| Instance name | nrp-2-0.5 | nrp-3-0.5 | nrp-4-0.5 | nrp-5-0.5 |
|---|---|---|---|---|
| Original scale | 500 | 500 | 750 | 1000 |
| Scale reduced by approximate backbone% | 88.4 | 88.6 | 79.9 | 60.8 |
| Scale reduced by soft backbone% | 11.2 | 7.6 | 7.2 | 32.4 |
| Sum of scale reduced% | 99.6 | 96.2 | 87.1 | 93.2 |

For all the four instances, less than 25% of the instance scales are removed after one pair of instance reductions; the scales of new instances are still too large for the solving algorithm. Rather than a single pair of instance reductions, multiple pairs can sufficiently reduce the instance scale. The instance scale gradually decreases while the number of instance reductions increases. The curves in Fig. 5 indicate that nearly all the instance reductions can reduce the scales of instances. After 12 pairs of instance reductions, only less than 15% of the scale for each instance is left, e.g., in nrp-3-0.5, only 19 customers are left after these multiple instance reductions. Moreover, Fig. 5 shows that 10 to 12 pairs of instance reductions can provide reasonable shrinkage for the scale of each instance.

To show the ability of the multilevel strategy for the instance reduction, we summarize the values of reduced scales in Table 4. For each instance, apart from the original instance scale, we show the scales reduced by all the approximate backbones, the scales reduced by all the soft backbones, and the sum of all the scales reduced by instances reductions. For example, with 12 pairs of instance reductions for nrp-3-0.5, 88.6% of the scale is reduced by the approximate backbone while 7.6% of the scale is reduced by the soft backbone. For the four instances in Table 4, the scale reduced by fixing the approximate backbone is larger than that by fixing the soft backbone. The scale reduced by fixing the soft backbone is between 7% and 33% while the one by fixing the approximate backbone is between 60% and 89%. Especially, in nrp-2-0.5, the left scale is 0.4%, i.e., 2 customers. Nearly the whole instance scale of nrp-2-0.5 is reduced in the multiple instance reductions.

Based on the analysis above, we conclude that multiple instance reductions can effectively reduce instance scales. Both of the two kinds of backbones work well in the instance reductions. The approximate backbone based instance reduction can reduce a large part of the instance scale while the soft backbone based instance reduction can enhance the reduction by the approximate backbone. Thus, we employ this multilevel strategy to design our algorithm, BMA.

## 4.2 Framework of BMA

In Algorithm 1, we present the details of our algorithm, BMA. The framework of BMA contains 3 phases: reduction, solving, and refinement.

In the reduction phase, the algorithm reduces the scale of an NRP instance by fixing the approximate backbone and the soft backbone. The approximate backbone is generated as the intersection of a certain number of local optimal solutions, which are obtained by a specified local search operator; the soft backbone is generated from an NRP instance after the instance reduction. In the solving phase, the local search operator in the reduction phase is employed to approximately solve the final small instance. In the refinement phase, the algorithm combines the approximate backbone, the soft backbone, and the current solution to the reduced instance together to construct a solution to the original instance. Either the reduction phase or the refinement phase is an iterative procedure, which reduces the instances or extends the solutions using a multilevel strategy. The actual number of levels in BMA depends on two input parameters, namely the maximum number of levels and the minimum scale of instances. Moreover, the other input parameter of BMA is the number of local optimal solutions in each level of the reduction phase. This parameter constrains the scale and the quality of the backbone. In Section 5.4.1, we will present an experiment on this parameter, i.e., the number of local optimal solutions.

**Algorithm 1. Backbone based Multilevel Algorithm**

**Input**: instance $\Pi_1$, local search operator $\Xi$,
maximum number $\alpha$ of levels,
minimum scale $\beta$ of instances,
number $\gamma$ of local optimal solutions
**Output**: solution $X_1$

**Phase I.** ***Reduction***
**for** $k = 1$ **to** $\alpha$ **do**
  **if** $|\Pi_k| > \beta$ **then**
    obtain a set $\Gamma_k$ of $\gamma$ local optimal solutions by $\Xi$ to $\Pi_k$;
    calculate $\xi_k^a$ = Approximate-Backbone($\Pi_k, \Gamma_k$);
    reduce instance, $\Pi_k{}'$ = Instance-Reduction($\Pi_k$, $\xi_k^a$);
    calculate $\xi_k^s$ = Soft-Backbone($\Pi_k{}', \phi$);
    reduce instance, $\Pi_{k+1}$ = Instance-Reduction($\Pi_k{}'$, $\xi_k^s$);
  **endif**
**endfor**
count the actual number $\delta$ of levels in Phase I;
**Phase II.** ***Solving***
obtain a local optimal solution $X_{\delta+1}$ to $\Pi_{\delta+1}$ by $\Xi$;
**Phase III.** ***Refinement***
**for** $k = \delta$ **to** 1 **do**
  refine solution $X_k = X_{k+1} \cup \xi_k^a \cup \xi_k^s$ ;
**endfor**

In Fig. 6, we illustrate the process of BMA with the instance presented in Fig. 3. For this instance, the algorithm employs two-level reductions and refinements. In the first level reduction (Fig. 6(a)), the local search operator obtains a set $\Gamma_1 = \{X_1^1, X_1^2, X_1^3\}$ of 3 local optimal solutions to the instance $\Pi_1$. Thus, the first level approximate backbone is $\xi_1^a = \{(1,1),(4,0)\}$, i.e., the customer $s_1$ is selected while the customer $s_4$ is not. Since the requirements for the customer $s_1$ are all satisfied in the release, all these requirements for $s_1$ can be reduced. By fixing the approximate backbone $\xi_1^a$, a new instance $\Pi_1{}'$ with 5 customers and 5 requirements is generated after the instance reduc-tion. Then no requirement is left for the customer $s_2$. Thus, the soft backbone is generated as $\xi_1^s = \{(2,1)\}$ and the instance is further reduced. Similarly, in the second level reduction (Fig. 6(b)), a set $\Gamma_2 = \{X_2^1, X_2^2, X_2^3\}$ of 3 local optimal solutions is obtained for the instance $\Pi_2$ with 4 customers and 5 requirements. Thus, the second level approximate backbone is $\xi_2^a = \{(6,1)\}$. By fixing $\xi_2^a$, a new instance $\Pi_2'$ with 3 customers and 2 requirements is generated. Then no requirement is left for the customer $s_7$. Thus, the soft backbone is generated as $\xi_2^s = \{(7,1)\}$ and a new instance $\Pi_3$ is generated as well. For the local search operator, the instance with 2 customers and 2 requirements is easy to solve (Fig. 6(c)). The solution is $X_3 = \{(3,1),(5,0)\}$. At last, under the inverted sequence of reductions, the algorithm combines the current solution, the approximate backbones, and the soft backbones together to construct the solution for each level (Fig. 6(d)). The final solution $X_1$ to the original instance $\Pi_1$ is formed within two-level reductions and refinements.

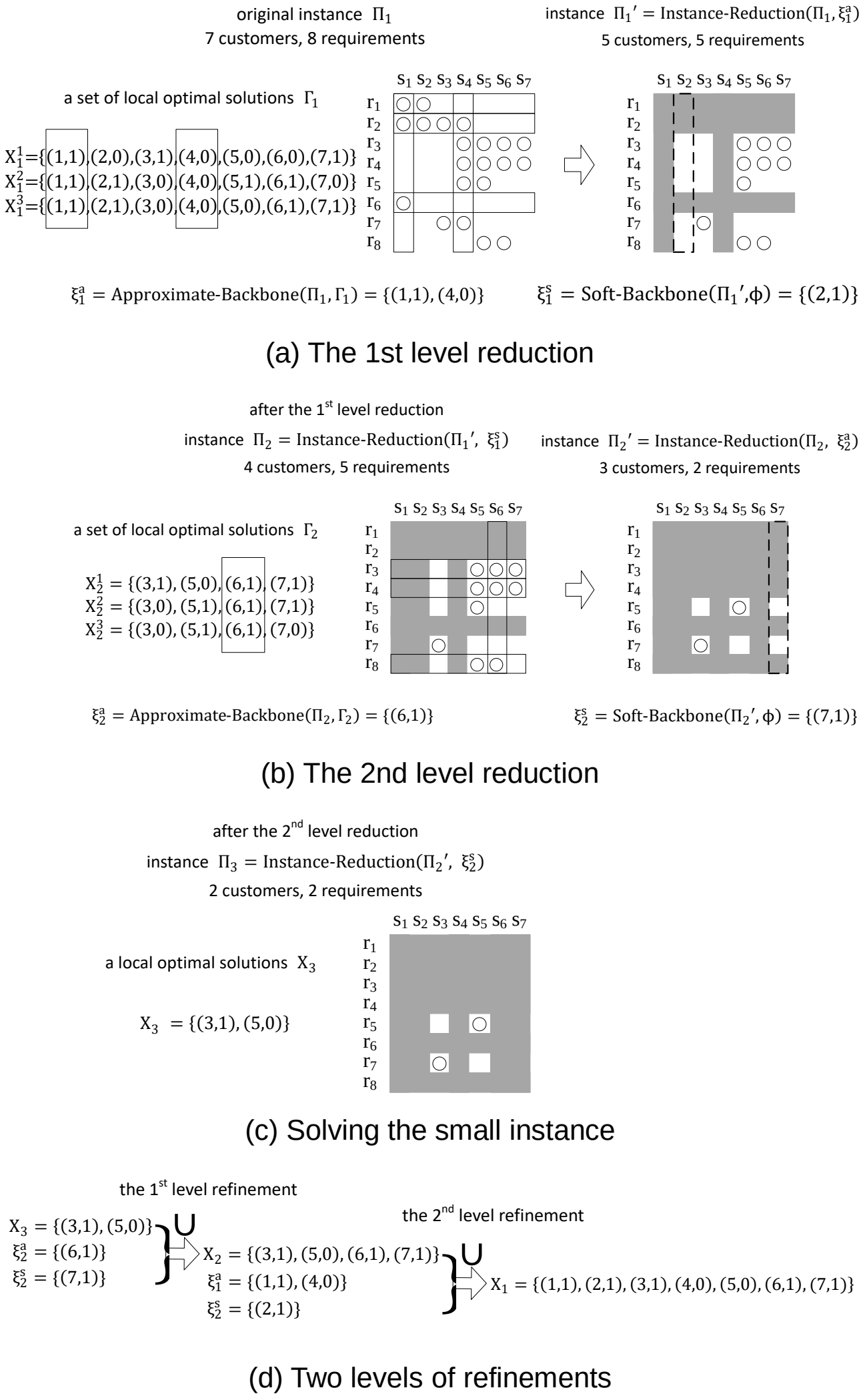

Fig. 6. Illustration on an instance with 7 customers and 8 requirements under two-level BMA. Sub-figures (a) and (b) show each level in the two-level reduction phase; (c) shows the solving phase; and (d) shows the two-level refinement phase.

## 5 Experiments and Results

For approximate algorithms, experimentation is a common way to evaluate the performance of algorithms. In this section, we evaluate our algorithm on 39 NRP instances. We first give the research questions in our experiments; then, we describe the instance generation rules of the classic NRP instances; next, we present the new instance generation method by mining open bug repositories; finally, we answer the research questions based on the experimental results.

### 5.1 Research Questions in the Experiments

We experimentally evaluate the performance of BMA for the NRP. For all the experiments in this paper, the algorithms are implemented with C++ and run on a PC with *Intel Core* 2.53 GHz processor and *uBuntu* OS (*Linux* kernel 2.6). We design the experiments to answer the following Research Questions (RQs):

**RQ1: Parameter configuration for BMA**. In the framework of BMA, each approximate backbone is generated based on a given number of local optimal solutions. The scale and the quality of the backbone (the combination of the approximate backbone and the soft backbone) may depend on the number of local optimal solutions, which is set manually for BMA. How does the number of local optimal solutions in BMA affect the backbone?

**RQ2: Performance evaluation**. In requirements engineering, some existing algorithms have been proposed to solve the NRP. We want to compare the solution quality of BMA with other algorithms. Can BMA perform well on the large scale NRP instances?

In Section 5.2 and Section 5.3, we will give the details of the NPR instances in our experiments. The NRP instances in this paper can be found in http://ssdut.dlut.edu.cn/oscar/nrp/.

### 5.2 Classic NRP Instance Generation

As requirements are usually private data of software companies, no open large NRP instances can be found in the literature. In this paper, we evaluate our algorithm on two sets of the NRP instances. One set includes 15 instances generated under certain constraints based on the classic literature of the NRP experiments [4]; the other set includes 24 realistic instances mined from open bug repositories of three open source software projects.

The classic set of the NRP instances consists of 5 groups and each group includes 3 instances. In each group, instances have distinct budget bounds, each of which equals to the *cost ratio* (0.3, 0.5, or 0.7, respectively) multiplied by the sum of all the costs. Table 5 shows the details of the 5 groups of instances. According to [4], these instances are based on Definition 1. Taken the group nrp-1 as an example, all the requirements are classified into 3 levels separated by the symbol "/". A requirement in the 2nd level may depend on some requirements in the 1st level while a requirement in the 3rd level may depend on some requirements in the 1st and 2nd levels. An instance name is formed by the group name and the cost ratio. For example, nrp-1-0.3 is an instance in the group nrp-1 and the cost ratio is 0.3. The details of the instance nrp-1-0.3 are as follows. There are 3 levels of requirements, 20, 40, and 80 requirements in each level. The costs of requirements in the three levels are from 1 to 5, from 2 to 8, and from 5 to 10, respectively. A requirement in the 1st level has at most 8 child requirements. Similarly, a requirement in the 2nd level has at most 2 child requirements. There are 100 customers, each of who requests 1 to 5 requirements. In addition, each customer provides a profit between 10 and 50.

TABLE 5
GENERATION RULES OF THE CLASSIC NRP INSTANCE GROUPS

| Instance group name | nrp-1 | nrp-2 | nrp-3 | nrp-4 | nrp-5 |
|---|---|---|---|---|---|
| # Requirements per level | 20/40/80 | 20/40/80/160/320 | 250/500/750 | 250/500/750/1000/750 | 500/500/500 |
| Cost of requirement | 1~5/2~8/5~10 | 1~5/2~7/3~9/4~10/5~15 | 1~5/2~8/5~10 | 1~5/2~7/3~9/4~10/5~15 | 1~3/2/3~5 |
| # Maximum child requirements | 8/2/0 | 8/6/4/2/0 | 8/2/0 | 8/6/4/2/0 | 4/4/0 |
| # Requests of customer | 1~5 | 1~5 | 1~5 | 1~5 | 1 |
| # Customers | 100 | 500 | 500 | 750 | 1000 |
| Profit of customer | 10~50 | 10~50 | 10~50 | 10~50 | 10~50 |

### 5.3 Mining Realistic NRP Instances

Besides the classic instances, we extract a set of NRP instances from open source bug repositories. To face the lack of large scale open requirements repositories, the requirements data can be mined from other databases. To our knowledge, only one requirements repository is mined for experiments, i.e., the requirements database mined from an open source forum project by Duan et al. [14]. In their paper, requests or problems in the forum project are mapped to the requests in a requirements repository to evaluate their requirements prioritization and triage approach.

In our work, to build large NRP instances, we mine the NRP instances from *bug repositories* (also called *bug tracking systems*, e.g., a popular bug repository, *Bugzilla* [7]). A bug repository is a database for the storage of numerous *bug reports*, each of which is submitted by a *user* (maybe a developer, a tester, or an end user) for recording the details of suggestions or problems. One bug report may be commented by one or more users; meanwhile, one user may make comments on one or more bug reports. The *user comments* on the bug reports provide a similar scenario for the requirements requests in requirements repositories. For example, if two users make comments on three bug reports in a bug repository, we can extract a software release, in which two customers request three requirements in the requirements analysis. Thus in our experiments, a bug report and a user in bug repositories are mapped to a requirement and a customer in the NRP, respectively. In addition, a user comment on a bug report is mapped to a requirement request; the *severity* of a bug report is mapped to the cost of a requirement. Similar to the classic set of NRP instances, the profit of a customer is randomly generated within a certain range. We present the corresponding relationship between bug repositories and the NRP in Table 6.

TABLE 6
CORRESPONDING RELATIONSHIP FOR THE ITEMS BETWEEN BUG REPOSITORIES AND THE NRP

| Item in the NRP | Item in a bug repository |
|---|---|
| Requirements, $R$ | Bug reports |
| Customers, $S$ | Users for the bug reports |
| Requests, $Q$ | User comments on bug reports |
| Costs, $C$ | The severity of the bug reports |
| Profits, $W$ | Random values generated within a certain range |

To mine the NRP instances, we employ the bug repositories in three open source software projects, namely *Eclipse* (a Java integrated development environment) [15], *Mozilla* (a set of web applications) [47], and *Gnome* (a desktop project) [19]. The XML form of these bug repositories can be found in *Mining Challenges 2007* and *2009* of *IEEE Working Conference on Mining Software Repositories (MSR)* [46]. To generate various instances, we set different parameters for bug repositories. In each group of instanc-

TABLE 7
DETAILS OF THE REALISTIC NRP INSTANCE GROUPS

| Instance group name | nrp-e1 | nrp-e2 | nrp-e3 | nrp-e4 | nrp-m1 | nrp-m2 | nrp-m3 | nrp-m4 | nrp-g1 | nrp-g2 | nrp-g3 | nrp-g4 |
|---|---|---|---|---|---|---|---|---|---|---|---|---|
| Source repository | Eclipse | | | | Mozilla | | | | Gnome | | | |
| Bug report ID | 150001~160000 | | 160001~170000 | | 200001~210000 | | 210001~220000 | | 450001~460000 | | 460001~470000 | |
| Bug reports time period | Jul. 2006~Oct. 2006 | | Oct. 2006~Jan. 2007 | | Mar. 2003~Jun. 2003 | | Jun. 2003~Sept. 2003 | | Jun. 2007~Jul. 2007 | | Jul. 2007~Aug. 2007 | |
| # Requests of customer | 4~20 | 5~30 | 4~15 | 5~20 | 4~20 | 5~30 | 4~15 | 5~20 | 4~20 | 5~30 | 4~15 | 5~20 |
| # Requirements | 3502 | 4254 | 2844 | 3186 | 4060 | 4368 | 3566 | 3643 | 2690 | 2650 | 2512 | 2246 |
| Cost of requirement | 1~7 | 1~7 | 1~7 | 1~7 | 1~7 | 1~7 | 1~7 | 1~7 | 1~7 | 1~7 | 1~7 | 1~7 |
| # Customers | 536 | 491 | 456 | 399 | 768 | 617 | 765 | 568 | 445 | 315 | 423 | 294 |
| Profit of customer | 10~50 | 10~50 | 10~50 | 10~50 | 10~50 | 10~50 | 10~50 | 10~50 | 10~50 | 10~50 | 10~50 | 10~50 |

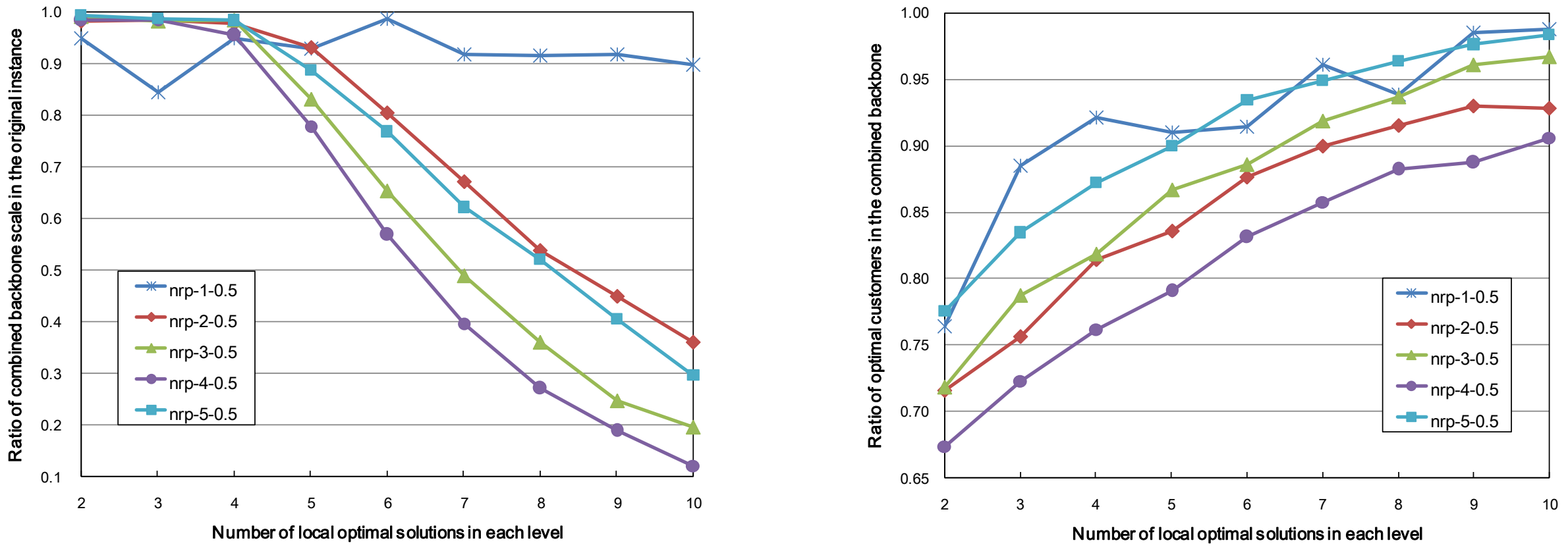

(a) Scale of the combined backbone

(b) Optimal customers in the combined backbone

Fig. 7. Relationship between the number of local optimal solutions, the scale of the combined backbone, and the ratio of optimal customers in the combined backbone.

es, first, we select 10000 continuous bug reports from a bug repository. The time period of the selected bug reports is around the software release time since the bug reports in this period are usually active [2]. Then, we filter out the bug reports and users (i.e., the requirements and customers in Table 6) out of a specified range by limiting the number of user comments (i.e., the requests in Table 6). As a result, the characteristics of a group can be generated. In Table 7, we show the details of 12 groups of instances extracted for experiments. The form of instances is based on Definition 2. Each group of instances consists of two instances, with the cost ratio 0.3 or 0.5, respectively. Therefore, the budget bound of each instance equals to the value of the sum of costs multiplied by the cost ratio. Thus, 24 realistic instances are mined for the following experiments.

### 5.4 Answers to Research Questions

In this section, we will answer the research questions proposed in Section 5.1. We evaluate our algorithm BMA on the 39 NRP instances mentioned in Sections 5.2 and 5.3.

#### *5.4.1. Answer to RQ1: Parameter configuration for BMA*

For the three input parameters of BMA, including the maximum number $\alpha$ of levels, the minimum scale $\beta$ of instances, and the number $\gamma$ of local optimal solutions, the parameters $\alpha$ and $\beta$ can be viewed as the termination conditions of BMA. However, the parameter $\gamma$ is a key value to decide the scale of the backbone. We experimentally evaluate the relationship among the number of local optimal solutions, the scale of the backbone, and the quality of the backbone.

In the framework of BMA, any algorithm can be embedded as a local search operator $\Xi$. To compare the experimental results, we employ the existing best local search algorithm, SA, to obtain local optimal solutions [7]. The solid empirical results and simplicity of SA have led to a wide range of applications in combinatorial optimization. In the experiments in this paper, we set the parameters for SA according to [4], i.e., the starting temperature is set to 100 and the non-linear ratio is set to $10^{-7}$.

In Fig. 7, we present the experimental results to visualize the relationship among the parameter $\gamma$, the scale of the backbone, and the quality of the backbone. To simplify the visualization, the backbone in Fig. 7 is a *combined backbone* (see Fig. 2), which is the combination of the approximate backbones and the soft backbones in all the levels of BMA. For a $\delta$-level BMA, given the approximate backbone $\xi_k^a$ and the soft backbone $\xi_k^s$ in the $k$th level, we define the combined backbone as $\xi^c = \bigcup_{1\le k\le\delta}(\xi_k^a \cup \xi_k^s)$.

We evaluate the scale and the quality of the combined backbone with two criteria, namely the ratio of the combined backbone scale and the ratio of optimal customers in the combined backbone. Given an NRP instance with the scale $n$, the ratio of the combined backbone scale in the original instance is $|\xi^c|/n$; given the best known solution $X^*$, the ratio of optimal customers in the combined backbone is $|\xi^c \cap X^*|/|\xi^c|$. In this experiment, we set $\alpha = 10$ and $\beta = 20$. Each point is calculated as an average of the results from ten independent runs.

In Fig. 7, the scale of the combined backbone decreases and the ratio of optimal customers increases while the number of local optimal solutions increases. Four of the curves in this experiment present the same trend when varying the number of local optimal solutions. The curve of the instance nrp-1-0.5 does not correspond with the curves of other instances since nrp-1-0.5 is a small instance, which is much easier to solve than the other four instances. Based on each value of $\gamma$, the instance scale of nrp-1-0.5 can be easily reduced. For all the five instances, when each approximate backbone in a level is generated by 2 local optimal solutions, the scale of the combined backbone is nearly the same as the instance scale and the number of optimal customers is less than 0.8 of the scale of the combined backbone; on the other hand, when each approximate backbone is generated by 10 local optimal solutions, the scale of the combined backbone is less than 0.4 of the instance scale for 4 out of 5 instances and the number of optimal customers is more than 0.9 of the scale of the combined backbone for all the 5 instances. From Fig. 7, we consider that 4 to 6 local optimal solutions for each approximate backbone is a good choice for the trade-off between the scale of the combined backbone and the number of optimal customers.

Based on this part, the answer to RQ1 is that the value of the input parameter $\gamma$, i.e., the number of local optimal solutions, can affect the scale and the quality of the backbone of BMA. The large scale backbone and the high quality backbone cannot be obtained simultaneously while tuning the value of $\gamma$. Thus, for the following experiments, we choose a trade-off value 5 for $\gamma$, which can balance the scale of the combined backbone and the number of optimal customers obtained by BMA. For other parameters $\alpha$ and $\beta$ in BMA, we choose parameter values as follows. In Fig. 5, we have studied the influence of the change of instance scales by tuning the pair of instance reductions. We set $\alpha = 10$ since over 10 pairs of instance reductions may significantly reduce the instance scale. For the parameter of the minimum scale of an instance, we manually set $\beta = 20$, since an instance with the scale less than 20 could be easy to solve [13].

### 5.4.2. Answer to RQ2: Performance evaluation

To evaluate the performance of BMA, we employ two direct solving algorithms for comparison. One algorithm is a Multi-Start strategy based SA (MSSA) [39]. In MSSA, the existing best local search algorithm, SA ([4], [5]) is run independently for multiple times and the best solution among these runs is chosen as the final solution [45]. The other algorithm is Genetic Algorithm (GA), which is a bio-inspired and population-based technology for complex problems, also for the NRP [69], [13]. Among many variants of GA, we choose the implementation described in [13]. This implementation uses an elitism based selection strategy to construct the population and updates new population with crossover and mutation operators.

We show the experimental results for the comparison among MSSA, GA and BMA on the NRP instances. SA is employed as a local search operator in both MSSA and BMA; the input parameters of SA are the same as those in Section 5.4.1. We set the parameters of MSSA and BMA as follows. In MSSA, we repeat SA for 30 times and choose the best solution; in BMA, we set the parameters according to Section 5.4.1, i.e., $\alpha = 10$, $\beta = 20$, and $\gamma = 5$.

To our knowledge, there is no prior parameter value of GA for the large scale NRP. Thus, we tune the parameters for GA. To this end, we configure the parameters for GA with an open access tuning tool, *ParamILS* [26], which employs an off-line local search framework for automatically tuning parameters. In *ParamILS*, we set the training set as three instances nrp-1-0.5, nrp-3-0.5, and nrp-5-0.5; we set the test set as the two instances nrp-2-0.5 and nrp-4-0.5. The cutoff time of *ParamILS* is set to 5000 seconds. After the parameter tuning by *ParamILS*, the values obtained are 100 for the population size, 0.3 for the elitism selection ratio, 0.3 for the crossover ratio, and 0.1 for the mutation ratio. Based on the parameters for GA, we set the maximum number of iterations to $10^5$. We choose such a value for the number of iterations to sufficiently show the solution quality of GA and to balance the running time of three algorithms.

We independently run each of the three algorithms (MSSA, GA, and BMA) for 10 times. The results are collected to measure the performance and to plot the profit distributions. In Table 8 and Table 9, we show the experimental results of MSSA, GA, and BMA on two sets of NRP instances. Each table has five columns, including the details of instances, the results of MSSA, the results of GA, the results of BMA, and the profit distributions. In the first column, the sub-columns are the instance name and the budget bound. The following three columns include sub-columns for the best profit, the average profit, and the average running time. For BMA, the sub-column "$MSSA\%$" and "$GA\%$" present the rate of average profit in percentage to measure the advantage by BMA against that by MSSA and GA. For example, "$MSSA\%$" is calculated as $(\omega_{BMA} - \omega_{MSSA})/\omega_{MSSA}$, where $\omega_{BMA}$ and $\omega_{MSSA}$ are the average profits obtained by BMA and MSSA, respectively. The average profit is used to measure the quality while the best profit is listed as a reference. Since each of the three algorithms is run for 10 times, we show the profit distributions of solutions with box plots [41] for all the algorithms in the last column. In a box plot, we measure the stability of solutions with the range between the first quartile and the third quartile. To normalize profits of distinct instances, the point in box plots is calculated as $\left(\omega(X) - \overline{\omega(X)}\right)/\overline{\omega(X)}$, where $\omega(X)$ is the profit obtained by the solution $X$ and $\overline{\omega(X)}$ is the average profit of all the solutions by an algorithm. Thus, the 0% point shows that the profit equals to the average. Note that based on the normalized profit distributions, each profit distribution shows the distribution for only one algorithm on one instance while no comparison is conducted for the absolute values of profits among MSSA, GA, and BMA.

In Table 8 for the classic instances, BMA obtains better solutions within less running time than MSSA and GA on most of the instances. Based on the sub-column "$MSSA\%$", the average profits obtained by BMA are 2% to 51% better than those by MSSA on all the instances. Note that on only two instances, the average profits by BMA are less than 10% better than those by MSSA, namely nrp-3-0.7 and nrp-5-0.7. The reason for this result is that the large cost ratio 0.7 makes it easy for MSSA to solve these instances, i.e., the predefined cost is adequate for making the decision. Thus, on these three instances, BMA can do only a little better than MSSA. On the other hand, based on the sub-column "$GA\%$", the average profits by BMA are 0% to 68% better than those by GA on all but one instance. The exception instance is nrp-1-0.5, on which GA can obtain better solutions than BMA. Moreover, on the other two instances in the group nrp-1 (with scale 100), the profits obtained by BMA are very similar to the profits by GA. That is, GA can work well on small scale instances. Among the rates in "$MSSA\%$" and "$GA\%$", both the rates less than 1% and the rates more than 60% are provided by "$GA\%$". As a result, we can find that the solutions of GA are in a wider range than those of MSSA. From the profit distributions of MSSA, GA, and BMA, the average profits on most of the instances are surrounded by the ranges of the profits. Moreover, among the last 9 instances (the last 3 groups) in Table 8, BMA can provide the most stable solution distributions for 6 instances (i.e., in box plots, for each of these 6 instances, the distance between the first quartile and the third quartile is short). In summary of Table 8, the results show that the backbone based instance reduction makes BMA obtain good solutions for the large scale NRP.

In Table 9 for the realistic instances, the experimental results are basically similar to those in Table 8. BMA can obtain the best solutions on all the instances. The average profits obtained by BMA are 19% to 35% better than those by MSSA on all the instances while the average profits obtained by BMA are 5% to 21% better than those by GA on all the instances. GA can beat MSSA on all these instances. From the sub-column "*GA%*", the rates on the instances extracted from Gnome (the instance names starting with "nrp-g") are smaller than the instances from Eclipse and Mozilla. In other words, the advantage of BMA is inconspicuous for the instances extracted from Gnome. A reason for this fact is that Gnome provides the simplest instances in our experiments, the instance scales of which are less than 500 (see Table 7 for the instance scales). On the contrary, for the large scale instances extracted from Mozilla, BMA can obtain much better profits than MSSA and GA. On the instances in Table 9, the profit distributions are also stable. On 10 instances among all the 24 realistic instances, the results obtained by BMA are the most stable in the three algorithms. From the sub-columns "*MSSA%*" and "*GA%*" in both Table 8 and Table 9, in general, the rate of profits decreases while the cost ratio increases for the instances in each group (i.e., from 0.3 to 0.7 for the classic instances or from 0.3 to 0.5 for the realistic instances). Thus, BMA can obtain much better solutions on most of the instances, which are with small cost ratios.

Based on this part, the answer to RQ2 is that BMA can obtain better solutions than the typical algorithms MSSA and GA on the large scale NRP instances. Moreover, the profit distributions provided by BMA are stable for most of the instances.

In conclusion of experiments in this section, the results show that BMA can obtain better profits than MSSA and GA within similar time on the large scale NRP instances. From the perspective of algorithm design, the approximate backbone leads to the fast solving for BMA; the soft backbone is helpful in constructing the near-optimal solutions to the problem instances; and the multilevel reductions and refinements provide a framework to use the existing algorithms. Based on these characteristics, BMA outperforms the typical algorithms, MSSA and GA, on most of the NRP instances.

## 6 Threats to Validity

Our approach is a search based technology to solve the NRP in requirements engineering. There are three potential threats to validity for our work.

### 6.1 Problem Definition

In this paper, only one kind of requirements dependency is given to the NRP model following the existing definitions [4], [5], [31]. However, there are some other kinds of dependencies in requirements engineering. For example, Carlshamre et al. [8] list six kinds of requirements dependencies and the dependency in our work can be viewed as a "REQUIRES" dependency in their approach; Zhang et al. [65] explore four kinds of requirements dependencies to facilitate the requirements reuse and software design.

Since the requirements dependencies in the NRP are formed as input parameters, it is straightforward to add other kinds of requirements dependencies to the model of the NRP. Based on the definition of the Simplified NRP, the model aims to handle the requirements requested by

TABLE 8
Performance for MSSA, GA, and BMA on 15 Classic Instances

| Instance | | MSSA | | | GA | | | BMA | | | | |
|---|---|---|---|---|---|---|---|---|---|---|---|---|
| Name | Bound | Best | Average | Time | Best | Average | Time | Best | Average | Time | MSSA% | GA% |
| nrp-1-0.3 | 257 | 998 | 976.5 | 108.65 | 1187 | 1178.1 | 85.63 | 1201 | 1188.3 | 52.68 | 21.69 | 0.87 |
| nrp-1-0.5 | 429 | 1536 | 1505.2 | 98.93 | 1820 | 1806.1 | 99.22 | 1824 | 1796.2 | 55.91 | 19.33 | -0.55 |
| nrp-1-0.7 | 600 | 2301 | 2273.6 | 91.70 | 2507 | 2505.4 | 79.19 | 2507 | 2507.0 | 34.51 | 10.27 | 0.06 |
| nrp-2-0.3 | 1514 | 3220 | 3158.3 | 320.76 | 2794 | 2737.0 | 654.23 | 4726 | 4605.6 | 246.14 | 45.83 | 68.27 |
| nrp-2-0.5 | 2524 | 5229 | 5094.1 | 288.70 | 5363 | 5276.4 | 891.55 | 7566 | 7414.1 | 280.87 | 45.54 | 40.51 |
| nrp-2-0.7 | 3534 | 8002 | 7922.6 | 255.52 | 9018 | 8881.1 | 911.55 | 10987 | 10924.7 | 277.47 | 37.89 | 23.01 |
| nrp-3-0.3 | 2661 | 5147 | 5088.8 | 461.21 | 5851 | 5719.0 | 910.99 | 7123 | 7086.3 | 436.90 | 39.25 | 23.91 |
| nrp-3-0.5 | 4435 | 8725 | 8553.4 | 420.66 | 9639 | 9574.2 | 542.22 | 10897 | 10787.2 | 438.80 | 26.12 | 12.67 |
| nrp-3-0.7 | 6209 | 13600 | 13518.2 | 489.79 | 12454 | 12360.7 | 265.23 | 14180 | 14159.2 | 215.90 | 4.74 | 14.55 |
| nrp-4-0.3 | 6648 | 6797 | 6708.4 | 1153.34 | 6675 | 6595.7 | 1849.15 | 9818 | 9710.5 | 854.48 | 44.75 | 47.22 |
| nrp-4-0.5 | 11081 | 11355 | 11120.6 | 1017.39 | 12781 | 12595.4 | 1587.22 | 15025 | 14815.5 | 907.03 | 33.23 | 17.63 |
| nrp-4-0.7 | 15513 | 19077 | 18830.1 | 1104.26 | 17327 | 17189.9 | 549.71 | 20853 | 20819.7 | 672.60 | 10.57 | 21.12 |
| nrp-5-0.3 | 1198 | 11421 | 11279.2 | 502.87 | 10689 | 10507.0 | 3069.26 | 17200 | 17026.9 | 475.85 | 50.96 | 62.05 |
| nrp-5-0.5 | 1996 | 17843 | 17756.6 | 472.48 | 18950 | 18732.9 | 1696.38 | 24240 | 24087.5 | 459.05 | 35.65 | 28.58 |
| nrp-5-0.7 | 2794 | 28347 | 28232.5 | 628.25 | 22174 | 22026.5 | 376.57 | 28909 | 28894.2 | 171.70 | 2.34 | 31.18 |

Profit distribution %

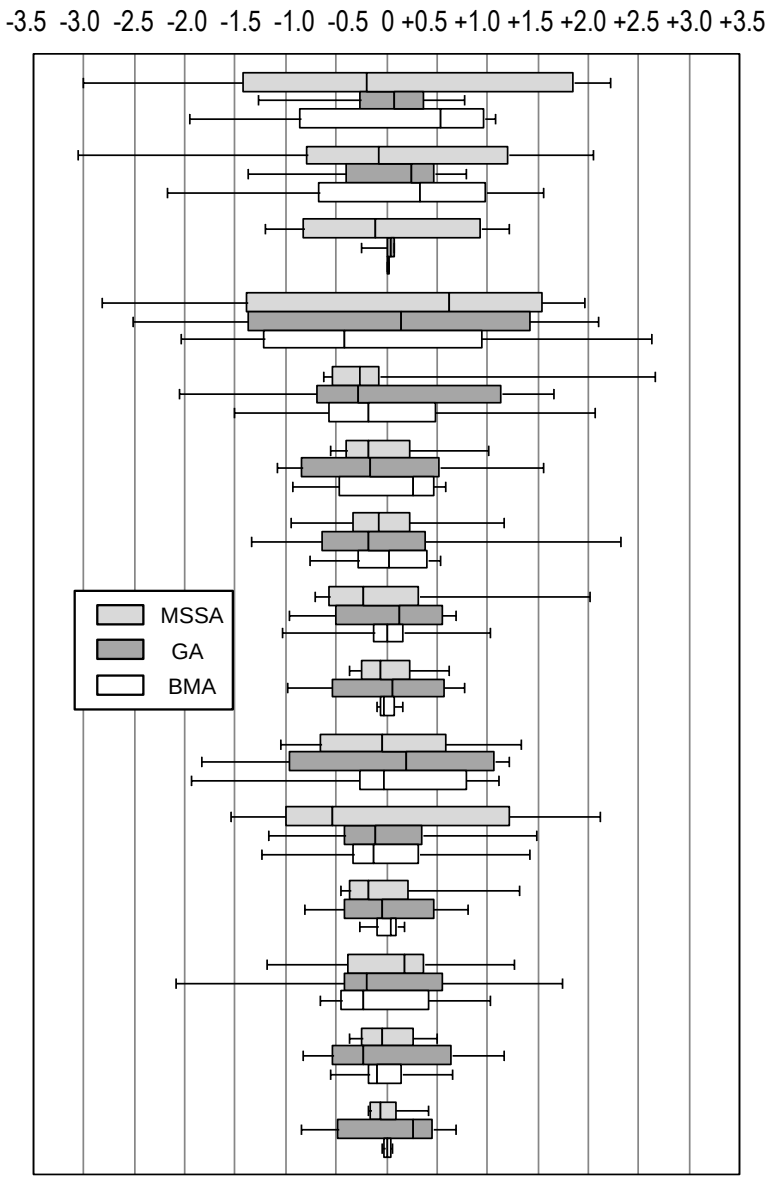

customers. As a result, the dependencies can be formed to the requirements requested by each customer. Therefore, our algorithm, BMA, can be extended to solve the NRP with various requirements dependencies.

### 6.2 Algorithm Construction

In this paper, we use the approximate backbone and the soft backbone to replace the backbone to construct our algorithm. The basic principle for using the approximate backbone is based on an empirical study, the fitness landscape analysis. However, it is not exact when applying the fitness landscape analysis for the relationship between the backbones and the approximate backbones. A theoretical analysis can provide much knowledge to the application of the approximate backbone. To our knowledge, the fitness landscape analysis is a useful empirical technology for approximately exploring the relationship between solutions [44], [35], [54]. This approximation between local optimal solutions and the optimal solutions can be viewed as a trade-off between theory and algorithm performance.

In the fitness landscape analysis, we use the best known solutions to replace the optimal solutions. This replacement may bring some perturbation to the analysis results. Since the optimal solutions to large scale instances are always hard to find within polynomial time, we follow the existing approaches to choose the most similar substitutions, i.e., the best known solutions [44], [54].

The soft backbone, another approximation of the backbone in our work, is also experimentally evaluated. Experimental results on four classic instances (in Fig. 5) have indicated the necessity of the soft backbone. However, an exact theoretical analysis is much better to quantify the power of the soft backbone, e.g., how to analyze the scale of the soft backbone for a given NRP instance. For both the approximate backbone and the soft backbone, a deep theoretical analysis may provide a further guideline for the design of backbone based algorithms.

### 6.3 Instance Bias

In the experimental results, we evaluate our algorithm on two sets of the NRP instances, namely a set generated under given constraints and a set extracted from bug repositories. However, both of these two sets of instances may bring threats to validity of our experimental results. On one hand, the classic generated NRP instances are a series of controllable randomized instances. Compared with real requirements repositories, these generated instances could provide extra stochastic distributions for the requirements data. On the other hand, the new ex-tracted instances are much realistic since the items in bug repositories can be viewed as a kind of requirements information. However, the knowledge gap between bug repositories and requirements repositories may lead to a bias for the evaluation results. To avoid the bias between our instances and real requirements data, the best method is to build open large requirements repositories in the future.

TABLE 9
PERFORMANCE FOR MSSA, GA, AND BMA ON 24 REALISTIC INSTANCES

| Instance | | MSSA | | | GA | | | BMA | | | | | Profit distribution % |
|---|---|---|---|---|---|---|---|---|---|---|---|---|---|
| Name | Bound | Best | Average | Time | Best | Average | Time | Best | Average | Time | MSSA% | GA% | |
| nrp-e1-0.3 | 3945 | 5723 | 5656.5 | 1181.73 | 6662 | 6553.4 | 1091.81 | 7572 | 7528.2 | 913.09 | 33.09 | 14.87 | |
| nrp-e1-0.5 | 6575 | 8550 | 8526.2 | 1103.49 | 9801 | 9756.3 | 715.17 | 10664 | 10589.2 | 927.05 | 24.20 | 8.54 | |
| nrp-e2-0.3 | 4722 | 5321 | 5281.0 | 1599.27 | 6275 | 6219.6 | 1076.41 | 7169 | 7109.9 | 1461.41 | 34.63 | 14.31 | |
| nrp-e2-0.5 | 7871 | 7932 | 7869.6 | 1522.26 | 9203 | 9172.9 | 778.67 | 10098 | 9999.8 | 1444.93 | 27.07 | 9.01 | |
| nrp-e3-0.3 | 4778 | 5031 | 4906.4 | 786.23 | 5795 | 5693.1 | 812.61 | 6461 | 6413.0 | 642.97 | 30.71 | 12.65 | |
| nrp-e3-0.5 | 7964 | 7436 | 7340.5 | 740.01 | 8491 | 8391.1 | 538.52 | 9175 | 9100.1 | 665.02 | 23.97 | 8.45 | |
| nrp-e4-0.3 | 5099 | 4332 | 4267.9 | 881.84 | 5065 | 5023.8 | 739.66 | 5692 | 5636.2 | 702.62 | 32.06 | 12.19 | |
| nrp-e4-0.5 | 8499 | 6459 | 6391.1 | 831.31 | 7487 | 7418.9 | 545.73 | 8043 | 7968.0 | 752.19 | 24.67 | 7.40 | |
| nrp-m1-0.3 | 3983 | 7733 | 7501.1 | 1636.15 | 8268 | 8188.3 | 1821.50 | 10008 | 9889.6 | 1445.38 | 31.84 | 20.78 | |
| nrp-m1-0.5 | 6639 | 11532 | 11450.1 | 1513.71 | 13287 | 13030.8 | 1226.71 | 14588 | 14437.7 | 1486.81 | 26.09 | 10.80 | |
| nrp-m2-0.3 | 3120 | 6321 | 6151.9 | 1729.67 | 6928 | 6863.9 | 1414.14 | 8272 | 8147.5 | 1607.23 | 32.44 | 18.70 | |
| nrp-m2-0.5 | 5200 | 9369 | 9265.3 | 1592.72 | 10873 | 10776.5 | 1044.93 | 11975 | 11883.5 | 1649.63 | 28.26 | 10.27 | |
| nrp-m3-0.3 | 3788 | 7427 | 7311.3 | 1501.49 | 8091 | 8016.1 | 1739.16 | 9559 | 9499.7 | 1081.69 | 29.93 | 18.51 | |
| nrp-m3-0.5 | 6313 | 11500 | 11380.9 | 1408.42 | 12969 | 12853.4 | 1099.90 | 14138 | 14036.6 | 1131.92 | 23.33 | 9.21 | |
| nrp-m4-0.3 | 3510 | 5719 | 5584.2 | 1281.95 | 6413 | 6341.3 | 1167.13 | 7408 | 7319.3 | 1054.64 | 31.07 | 15.42 | |
| nrp-m4-0.5 | 5850 | 8770 | 8641.5 | 1191.33 | 9970 | 9923.2 | 827.70 | 10893 | 10790.7 | 1089.39 | 24.87 | 8.74 | |
| nrp-g1-0.3 | 4140 | 4806 | 4723.0 | 729.61 | 5494 | 5437.0 | 733.69 | 5938 | 5911.3 | 614.23 | 25.16 | 8.72 | |
| nrp-g1-0.5 | 6900 | 7339 | 7250.3 | 692.65 | 8223 | 8151.7 | 468.43 | 8714 | 8660.0 | 625.74 | 19.44 | 6.24 | |
| nrp-g2-0.3 | 3677 | 3583 | 3549.9 | 781.94 | 4256 | 4195.5 | 523.65 | 4526 | 4486.2 | 576.97 | 26.38 | 6.93 | |
| nrp-g2-0.5 | 6129 | 5433 | 5359.8 | 758.83 | 6219 | 6138.4 | 398.52 | 6502 | 6470.2 | 566.72 | 20.72 | 5.41 | |
| nrp-g3-0.3 | 4258 | 4663 | 4593.1 | 659.25 | 5351 | 5296.6 | 665.49 | 5802 | 5736.5 | 563.12 | 24.89 | 8.31 | |
| nrp-g3-0.5 | 7097 | 7032 | 6948.1 | 625.68 | 7903 | 7849.8 | 425.30 | 8402 | 8326.8 | 576.81 | 19.84 | 6.08 | |
| nrp-g4-0.3 | 3210 | 3386 | 3313.9 | 658.86 | 3951 | 3909.9 | 444.16 | 4190 | 4159.0 | 475.27 | 25.50 | 6.37 | |
| nrp-g4-0.5 | 5350 | 5041 | 4991.6 | 626.72 | 5751 | 5721.3 | 335.42 | 6030 | 5986.5 | 473.58 | 19.93 | 4.64 | |

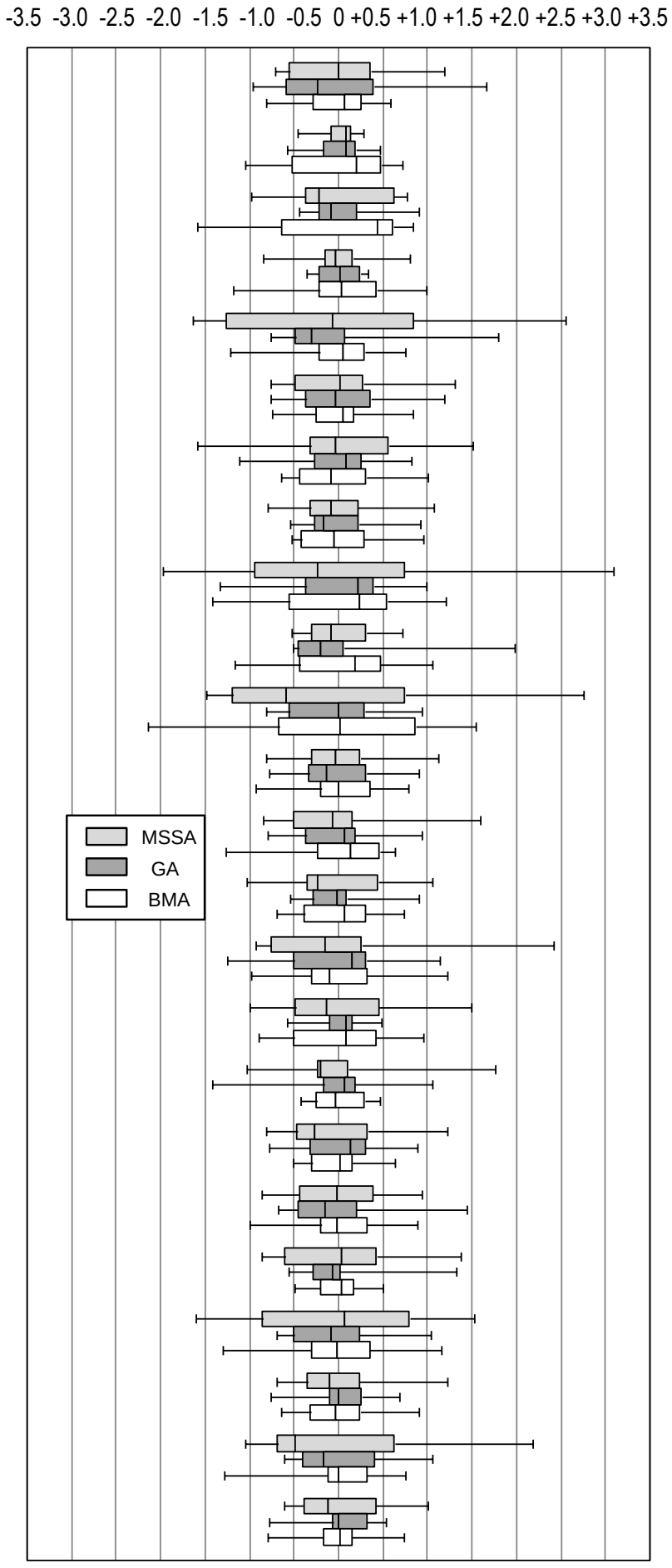

## 7 Related Work

To our knowledge, this paper is the first work using backbone based algorithms to solve requirements engineering problems. In this section, we investigate the related work of this paper.

### 7.1 The NRP and Requirements Selection

To balance customer profits and requirements costs, Bagnall et al. [4] first proposed the NRP in 2001. In this work, they model the NRP, provide the instance generation rules, and apply numerous search based algorithms to approximately solve the NRP. The most relevant problem of the NRP is the process of Release Planning (RP) [21], which addresses selecting optimal releases to satisfy software requirements constraints [56] or release time [62], [40]. Both the NRP and the RP aim to find an optimal decision for requirements selection, especially dependency constraints based requirements selection. The NRP tends to address customer profits for the coming release while the RP tends to directly assign requirements for multiple releases. A recent review of the RP by Svahnberg et al. [59] lists and compares the related work of the RP.

Based on the number of problem objectives, the related work of the NRP can be divided into two categories, namely single-objective and multi-objective. In the single-objective NRP (or the NRP for short), such as the problem in this paper, the cost bound of a software release is predefined and the problem objective is to obtain the maximum profits from customers. For example, Geer & Ruhe [21] propose a genetic algorithm based approach to optimize software releases; Jiang et al. [31] propose an ant colony optimization algorithm to solve the NRP; Baker et al. [5] extend the NRP with component prioritization and solve this problem with the greedy algorithm and the simulated annealing. Moreover, for the resource allocation for software releases, Ngo-The & Ruhe [50] propose a two-phase optimization by combining integer programming to relax the search space and genetic programming to reduce the search space. In this paper, we address the large scale single-objective NRP. Our approach is to downgrade the problem scale in contrast to the existing algorithms, which solve the problems directly.

In the Multi-Objective NRP (MONRP), besides the profit, another objective is usually defined to minimize software costs. Zhang et al. [69] first proposed the MONRP and gave an empirical study with the genetic algorithm based multi-objective optimization algorithms in 2007. Many extensions of the MONRP are studied to balance the benefits and resources, including fairness [17], sensitivity [25], and robustness with completion time [22]. Moreover, Saliu & Ruhe [57] detect feature coupling from both business perspectives and implementation perspectives; Zhang et al. [66] model two periods of profits to analyze the requirements under varying time. A recent work by Zhang & Harman [68] shows that the dependencies in requirements interaction management can be formulated as an extension of the MONRP.

The NRP is a combinatorial optimization model for requirements selection. Requirements selection and optimization have impacted numerous aspects of requirements engineering, including requirements management [55], [48], [49], requirements prioritization [32], [3], requirements triage [12], [14], and requirements visualization [16]. In addition, for further researches in requirements selection, some work investigates requirements interdependencies to explore the relationship and conflicts between requirements [8], [65], [20], [68].

### 7.2 Search Based Requirements Engineering

By fixing the optimal requirements for the next release, our work is a kind of Search Based Software Engineering (SBSE) approach for requirements engineering. In SBSE, software engineering problems are transformed into optimization problems for approximately solving with search technologies [24], [23]. One typical field of SBSE is search based software testing (e.g., [42], [36], [1], [43]). Some other fields of SBSE include design (e.g., [6], [53], [61]), quality (e.g., [37]), refactoring (e.g., [51]), reverse engineering (e.g., [34]), etc.

Among the fields of SBSE, Search Based Requirements Engineering (SBRE) aims to manage requirements with search technologies [24]. Most of work about the NRP and its relevant problems is the typical application of SBRE. A survey of SBRE shows the existing work and challenges in this field [67]. In our work, the backbone based algorithm is introduced to SBRE for the first time.

### 7.3 Backbone and its Application

The backbone, a basic structure for reductions and refinements in our work, is a solving strategy for exploring the hard problems in combinatorial optimization [58], [64], [33], [30]. To our knowledge, besides our work, there is only one concept similar to the approximate backbone for search technologies in software engineering. That is, Mahdavi et al. [38] propose a *building block* based multiple hill climbing approach for the software module clustering problem. From the viewpoint of combinatorial optimization, a building block in [38] is also an intersection of local optimal solutions as the approximate backbone. However, in our work, the concept of the soft backbone is first proposed in both software engineering and combinatorial optimization. The soft backbone and the multilevel strategy are combined with the approximate backbone to solve large scale search based problems.

Besides the backbone, the *muscle* and the *fat* in combinatorial optimization are two other effective technologies for constructing the solutions. The muscle of an instance is the union set of optimal solutions [29], [27] while the fat of an instance is the part without any optimal solution [11]. Drawn on the experiments from the existing work in combinatorial optimization, each of the backbone, the muscle, and the fat can be employed to further guide the algorithm design for problems in requirements engineering.

### 7.4 Large Scale Optimization

The multilevel approach is a kind of large scale optimization technology [60]. As we mentioned in Section 4.1, the key idea of the multilevel approach is to iteratively convert the original problem into multiple sub-problems so that the algorithm can downgrade the problem scale to apply existing algorithms. In this paper, our BMA is a multilevel approach for reducing the problem scale in requirements engineering.

Besides the multilevel approach, the cooperative co-evolution approach is one of the recently proposed technologies for large scale optimization [63], [52]. In contrast to the iterative reduction in the multilevel approach, the cooperative co-evolution approach employs the divide-and-conquer strategy to find the optimal solutions.

## 8 Conclusions and Future Work

As an important problem in requirements engineering, the Next Release Problem (NRP) aims to balance customer profits and requirements costs for the project decision. In this paper, we propose a Backbone based Multilevel Algorithm (BMA) to solve the large scale NRP. Based on the approximate backbone and the soft backbone, BMA iteratively reduces the instance scales and refines the solutions to construct the final solution. Experimental results show that BMA can achieve better performance than direct solving approaches. In our work, we propose the soft backbone for the first time, which can be generated from the instance after the instance reduction. Moreover, we also propose a method to generate requirements data from open bug repositories. This method can be used to supplement the lack of open requirements databases.

Our future work will focus on the application of BMA to other problems in software engineering. In requirements engineering, BMA can be used to solve many other large scale problems, such as release planning and requirements prioritization. The backbone based multilevel strategy can build a bridge between large scale problems and existing algorithms. We will explore some new problems, which may be solved by the similar strategy of BMA. In addition, we plan to develop BMA with a theoretical analysis, e.g., how to estimate the scale of the backbone without empirical methods. Apart from applications in requirements engineering, we want to apply the BMA to the regression test case selection problem in software testing. The model of the regression test case selection problem is very similar to the NRP. Thus, the application of BMA can be extended to various fields in software engineering.

Another further work is to explore the relationship between open bug repositories and requirements repositories. In this paper, we map items in the NRP to ones in open bug repositories. However, the domain knowledge behind these two kinds of repositories may bring a gap to the application from one repository to the other. We will conduct an empirical study to find out the details of this knowledge gap.

## Acknowledgments

We greatly thank our anonymous reviewers for their insightful comments and corrections. We thank Edward Prendergast and Qingna Fan with Intel Corporation for their helpful suggestions.

This work is partially supported by the National Natural Science Foundation of China under grants 60805024 and 61033012, and the National Research Foundation for the Doctoral Program of Higher Education of China under grant 20070141020.

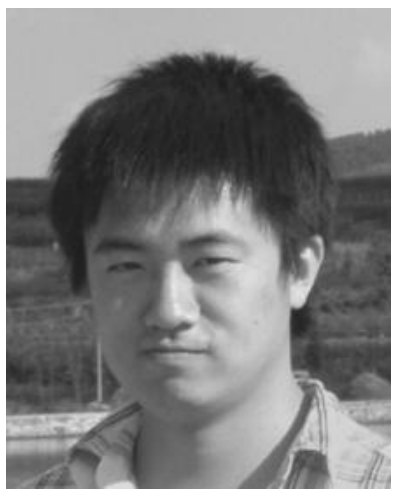

**Jifeng Xuan** received the BSc degree in software engineering from Dalian University of Technology, China, in 2007. He is currently working toward the PhD degree at Dalian University of Technology. His research interests include search based software engineering, mining software repositories, and machine learning. He is a student member of the China Computer Federation (CCF).

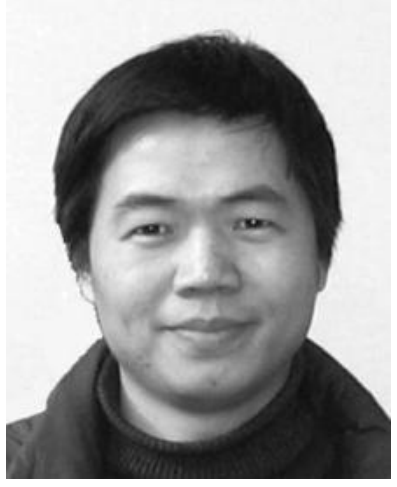

**He Jiang** received the BSc and PhD degrees in computer science from University of Science and Technology of China (USTC), China, in 1999 and 2005, respectively. He is currently an associate professor at School of Software, Dalian University of Technology, China. His research interests include computational intelligence and its applications in software engineering and data mining. He is a member of the IEEE and the China Computer Federation (CCF).

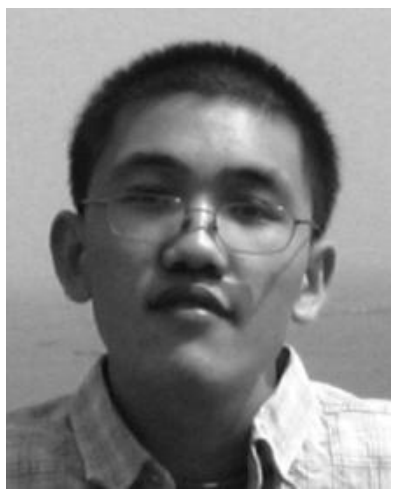

**Zhilei Ren** received the BSc degree in software engineering from Dalian University of Technology, China, in 2007. He is currently working toward the PhD degree at Dalian University of Technology. His research interests include metaheuristic algorithm design, data mining, and their applications in software engineering. He is a student member of the China Computer Federation (CCF).

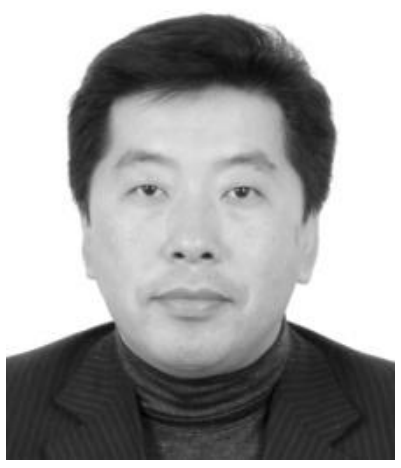

**Zhongxuan Luo** received the BSc degree in computational mathematics from Jilin University, China, in 1985, the MSc degree in computational mathematics from Jilin University, in 1988, and the PhD degree in computational mathematics from Dalian University of Technology, China, in 1991. He has been a full professor of School of Mathematical Sciences at Dalian University of Technology since 1997. His research interests include multivariate approximation theory and computational geometry.